\documentclass[preprint,12pt]{elsarticle}
\usepackage[all]{xy}
\usepackage{amssymb, amsmath}
\usepackage{accents}
\usepackage{pgfplots}
\pgfplotsset{compat=1.11}
\usepackage[title]{appendix}

\def\Xint#1{\mathchoice
   {\XXint\displaystyle\textstyle{#1}}%
   {\XXint\textstyle\scriptstyle{#1}}%
   {\XXint\scriptstyle\scriptscriptstyle{#1}}%
   {\XXint\scriptscriptstyle\scriptscriptstyle{#1}}%
   \!\int}
\def\XXint#1#2#3{{\setbox0=\hbox{$#1{#2#3}{\int}$}
     \vcenter{\hbox{$#2#3$}}\kern-.5\wd0}}

\def\dashint{\Xint-}

\newtheorem{lemma}{Lemma}
\newtheorem{corollary}[lemma]{Corollary}
\newtheorem{prop}[lemma]{Proposition}

\usepackage{tikz}
\usepackage{graphicx}
\usepackage{subcaption}
\usepackage{float}

\newcommand{\mi}{\mathrm{i}}

\journal{}

\begin{document}

\begin{frontmatter}

\title{Ground-state energies of the open and closed $p+ip$-pairing models from the Bethe Ansatz}
\author{Yibing Shen, Phillip S. Isaac, Jon Links$^*$\\ }
\address{School of Mathematics and Physics, \\The University of Queensland, Brisbane, QLD 4072, 
Australia \\
$^*$email: jrl@maths.uq.edu.au}
\begin{abstract}
Using the exact Bethe Ansatz solution, we investigate methods for calculating the  
ground-state energy for the $p + ip$-pairing Hamiltonian. 
We first consider the Hamiltonian isolated from its environment
(closed model) through two forms of Bethe Ansatz solutions, which generally have complex-valued
Bethe roots. A continuum limit approximation, leading to an integral equation, is
applied to compute the ground-state energy. We discuss the evolution of the
root distribution curve with respect to a range of parameters, and the limitations of
this method. We then consider an alternative approach
that transforms the Bethe Ansatz equations to an equivalent form, but in terms of the real-valued conserved operator eigenvalues. An integral equation is established for the transformed solution. This equation is shown to admit an exact solution associated with the ground state.
Next we discuss results for a recently derived Bethe Ansatz solution of the open model. With the
aforementioned alternative approach based on real-valued roots, combined with mean-field analysis, we
are able to establish an integral equation with an exact solution that
corresponds to the ground-state for this case.
\end{abstract}



\end{frontmatter}

\newpage 
\section{Introduction}
The $p+ip$-pairing Hamiltonian is an example of a Bardeen-Cooper-Schrieffer (BCS) model which admits an exact Bethe Ansatz solution. This result was initially established for the {\it closed system} which conserves particle number \cite{ilsz09,s09,dilsz10,rdo10,vdv14,lmm15}.
For the {\it open} model there is no conservation of total particle number, due to interaction terms which accommodate particle exchange with the system's environment. Consequently a $u(1)$ symmetry is broken, which generally renders the analysis of the system to be more complicated. Integrability of the open model was established in \cite{lil16} through use of the Boundary Quantum Inverse Scattering Method. An alternative derivation, which is less technical, was later provided in \cite{l17}. Topological properties of the open model, relating to zero-energy excitations, have been studied in \cite{cdv16}.

Like its better known ancestor the Richardson model \cite{r63}, which is associated with $s$-wave pairing, the existence of the exact solution for the closed $p+ip$ system provides a means to calculate the ground-state energy through a continuum limit approximation. Numerical studies suggest that, in the limit of infinite particle number, the ground-state roots become dense and lie on a connected curve in the complex plane. For Richardson's solution at half-filling this approach was first investigated by Gaudin \cite{g95}, and subsequently re-examined by Rom\'an et al. \cite{rsd02}. One way to view this problem is to use the language of a two-dimensional electrostatic analogy. It was found in this manner that the solution for the ground-state energy coincides with the prediction coming from mean-field calculations. It is worthy of  mention  that an extended discussion and application of the electrostatic analogy for more general Richardson-Gaudin systems can be found in \cite{admor02}. 

Following this approach, the continuum limit approximation has also been adopted in \cite{dilsz10,rdo10} for the closed $p+ip$-pairing model. Despite having much more complicated patterns of Bethe root distribution compared to the Richardson model,  the results were again found to give agreement with mean-field analysis. However, there are some technical issues concerning the assumptions made regarding the Bethe root distributions which warrant closer scrutiny. This is the first of the primary objectives of the current study. In particular a specific example will be provided which establishes that, for certain model parameters, the continuum limit approach fails to provide fully consistent equations describing the Bethe root behaviour. However, a surprising outcome is that although the method of calculation is flawed in some instances, the results remain valid. That is, in the continuum limit the ground-state energy is the same as that predicted by mean-field theory, across all values of the model parameters. This will be proved by exploiting a completely different approach which does not use the Bethe root distribution at all.

The origins of the new approach that will be followed trace back to the work of Babelon and Talalaev \cite{bt07} who showed that, through a change of variables, the Bethe Ansatz equations for Richardson-Gaudin type systems could be recast into a set of coupled polynomial equations. The roots of these equations are related to the eigenvalues of the self-adjoint conserved operators, and as such that are necessarily real-valued.  The same form of polynomial equations were adopted in \cite{fesg11} as a means of efficient numerical solution of the conserved operator spectrum, and in \cite{fs12} to compute wavefunction overlaps. In these instances the equations are quadratic. Extensions were given in \cite{cdvv15,cvd17} to a setting suitable for the $p+ip$ Hamiltonian, for which the polynomial equations are also quadratic. Here it will be shown that in this form, the continuum limit approach can be formulated and solved in such a way that it does not require an Ansatz for the distribution of the roots of the equations.       

The second primary objective is to apply this methodology for the calculation of the ground-state energy in the case of the {\it open} $p+ip$ Hamiltonian.  Here it will be shown how the alternative method developed to compute the ground-state energy in the closed case easily extends to the open case. It will also be shown that the result is again in complete agreement with mean-field calculations.   

The general form of the integrable Hamiltonian is introduced in Sect. \ref{sec:hamiltonian}. The closed model, for which the coupling constant of the environment interaction is set to zero, is then described in detail in Sect. \ref{sec:ppip0}. Two forms of Bethe Ansatz solution are presented, and the continuum limit approximation for calculating the ground-state energy is formulated. Following from this an analysis exposing the limitations of the continuum limit approximation is conducted. In Sect. \ref{sec:coem}, attention turns towards formulating an alternative approach, based on the set of transformed Bethe Ansatz equations which result in coupled quadratic equations. 
Then in Sect. \ref{sec:ppip1} the process is extended to accommodate the open model. Concluding remarks and discussion are offered in Sect. \ref{sec:conclusion}.

\section{The Hamiltonian} \label{sec:hamiltonian}

The annihilation and creation operators for two-dimensional fermions of mass $m$ with momentum $\mathbf{k} = k_x+ {\rm i} k_y$ are denoted by $c_\mathbf{k}$, $c_\mathbf{k}^\dagger$, satisfying $$\{ c_\mathbf{k},c_{\mathbf{k}'} \} = \{ c_\mathbf{k}^\dagger, c_{\mathbf{k}'}^\dagger \} =0, \quad \{ c_\mathbf{k}, c_{\mathbf{k}'}^\dagger \} = \delta_{\mathbf{k}\mathbf{k}'} I.$$
We consider the following Hamiltonian of the pairing model interacting with its environment \cite{cdv16}
\begin{align} \label{eq:ppip10}
    \mathcal{H} & = \sum_\mathbf{k} \frac{|\mathbf{k}|^2}{2m} c_\mathbf{k}^\dagger c_\mathbf{k} - \frac{\mathcal{G}}{4m} \sum_{\mathbf{k} \neq \pm \mathbf{k}'} \mathbf{k} \bar{\mathbf{k}'}c_\mathbf{k}^\dagger c_{-\mathbf{k}}^\dagger c_{-\mathbf{k}'} c_{\mathbf{k}'} + \frac{\Gamma}{2} \sum_\mathbf{k} ( \mathbf{k}c_\mathbf{k}^\dagger c_{-\mathbf{k}}^\dagger + \bar{\mathbf{k}}c_{-\mathbf{k}} c_{\mathbf{k}} ),
\end{align}
where $\mathcal{G}$, $\Gamma$ are positive real constants. The sum of momenta is taken over an index set $K$ with the properties (i) if ${\mathbf  k} \in K$, then $-{\mathbf  k} \in K$; (ii) for all ${\mathbf  k} \in K$ we have $|{\mathbf k}| \leq \omega$, where $\omega$ is called the {\it cut-off}.  The cardinality of $K$ is denoted as $2L$. The following equality is satisfied on this Hilbert subspace
\begin{equation*} 
    2c_\mathbf{k}^\dagger c_\mathbf{k} c_{-\mathbf{k}}^\dagger c_{-\mathbf{k}} = c_\mathbf{k}^\dagger c_\mathbf{k} + c_{-\mathbf{k}}^\dagger c_{-\mathbf{k}}.
\end{equation*}
Let $k_x + {\rm i}k_y=|\mathbf{k}| \exp({\rm i} \phi_\mathbf{k})$, we then introduce the following notation:
$$ S_\mathbf{k}^+ = \exp{(i\phi_\mathbf{k})}c_\mathbf{k}^\dagger c_{-\mathbf{k}}^\dagger , \quad S_\mathbf{k}^- = \exp{(-i\phi_\mathbf{k})}c_{-\mathbf{k}} c_{\mathbf{k}}, \quad S_\mathbf{k}^z = c_\mathbf{k}^\dagger c_{-\mathbf{k}}^\dagger c_{-\mathbf{k}} c_{\mathbf{k}} - \frac{I}{2}. $$
These operators satisfy the $\mathfrak{su}(2)$ algebra commutation relations
$$ [S_\mathbf{k}^z,S_\mathbf{k}^\pm] = \pm S_\mathbf{k}^\pm,\quad [S_\mathbf{k}^+, S_\mathbf{k}^-] = 2S_\mathbf{k}^z .$$
From now on, we use integers to enumerate the pairs of momentum states $\mathbf{k}$ and $-\mathbf{k}$. Setting the mass to be $m=1$ and $z_\mathbf{k}=|\mathbf{k}|$, we rewrite \eqref{eq:ppip10} as
\begin{align} \label{eq:ppip1}
   \mathcal{H} & = \sum_{k=1}^L z_k^2 S_k^z - \mathcal{G} \sum_{k=1}^L \sum_{j \neq k}^L z_k z_j S_k^+ S_j^- + \Gamma  \sum_{k=1}^L z_k (S_k^+ + S_k^-) + \left(\frac{1}{2}\sum_{k=1}^L z_k^2 \right)I.
\end{align}
Defining the following operators
\begin{align} \nonumber
    \mathcal{T}_j = \sum_{k \neq j}^L  & \left( \frac{z_k^2}{z_k^2-z_j^2}(4S_j^z S_k^z - I) + \frac{2z_j z_k}{z_k^2-z_j^2} (S_j^+ S_k^- + S_j^- S_k^+) \right)\\ \label{eq:tj1}
    & + \frac{2}{\mathcal{G}} S_j^z + \frac{2\Gamma}{\mathcal{G}} z_j^{-1} (S_j^+ + S_j^-),
\end{align}
it can be shown that
\begin{equation} \label{eq:hamtj}
\mathcal{H} = \frac{\mathcal{G}}{2} \sum_{j=1}^L z_j^2 \mathcal{T}_j + 
\left( \frac{1}{2}\sum_{j=1}^L z_j^2 \right) I,
\end{equation}
and $\{ \mathcal{T}_j \}$ is a set of mutually commuting conserved operators. These operators have been shown \cite{cdv16} to satisfy the following quadratic identities
\begin{equation} \label{eq:tj2}
    \mathcal{T}_j^2 = \frac{1}{\mathcal{G}^2} + \frac{4\Gamma^2 z_j^{-2}}{\mathcal{G}^2} + 2\sum_{k \neq j}^L \frac{z_k^2(\mathcal{T}_j-\mathcal{T}_k)}{z_j^2-z_k^2}.
\end{equation}
Hence the eigenvalues $\{\mathfrak{t}_j\}$ corresponding to $\{\mathcal{T}_j\}$ give the energy expression 
\begin{equation} \label{eq:egytj}
    \mathcal{E} = \frac{\mathcal{G}}{2} \sum_{j=1}^L z_j^2 \mathfrak{t}_j + \frac{1}{2}\sum_{j=1}^L z_j^2 ,
\end{equation}
and due to \eqref{eq:tj2} $\{\mathfrak{t}_j\}$ satisfy
\begin{equation} \label{eq:tj3}
    \mathfrak{t}_j^2 = \frac{1}{\mathcal{G}^2} + \frac{4\Gamma^2 z_j^{-2}}{\mathcal{G}^2} + 2\sum_{k \neq j}^L \frac{z_k^2(\mathfrak{t}_j-\mathfrak{t}_k)}{z_j^2-z_k^2}.
\end{equation}
\section{The $p+ip$ model isolated from the environment} \label{sec:ppip0}

The $p+ip$ Hamiltonian is isolated from the environment (closed model) when $\Gamma =0$. In this case, we adopt the letters $G, E$ and $T$ instead of $\mathcal{G}, \mathcal{E}$ and $\mathcal{T}$. The Hamiltonian reads
\begin{align} \label{eq:ppip0}
    H_0 & = \sum_{k=1}^L z_k^2 S_k^z - G \sum_{k=1}^L \sum_{j \neq k}^L z_k z_j S_k^+ S_j^-  + \left( \frac{1}{2}\sum_{k=1}^L z_k^2 \right) I\\ \nonumber
    & = \frac{G}{2} \sum_{j=1}^L z_j^2 T_j + \left( \frac{1}{2}\sum_{j=1}^L z_j^2 \right) I,
\end{align}
where 
\begin{align} \label{eq:tj1gamma0}
    T_j & = \sum_{k \neq j}^L \left( \frac{z_k^2}{z_k^2-z_j^2}(4S_j^z S_k^z - I) + \frac{2z_j z_k}{z_k^2-z_j^2} (S_j^+ S_k^- + S_j^- S_k^+) \right) + \frac{2}{G} S_j^z.
\end{align}
The quadratic identity \eqref{eq:tj2} in this case becomes
\begin{align} \label{eq:tj2gamma0}
    T_j^2 = \frac{1}{G^2} + 2\sum_{k \neq j}^L \frac{z_k^2(T_j-T_k)}{z_j^2-z_k^2}.
\end{align}
The eigenvalues $\{t_j\}$ of $\{T_j\}$ in \eqref{eq:tj1gamma0} then according to \eqref{eq:tj2gamma0} satisfy
\begin{equation} \label{eq:ppip0tj1}
    t_j^2 = \frac{1}{G^2} + 2 \sum_{k \neq j}^L \frac{z_k^2 (t_j - t_k)}{z_j^2 - z_k^2}, \quad j=1,2,\dots,L.
\end{equation}
Also note that in this case we have $$\sum_{j=1}^L T_j = 2\left( \sum_{k=1}^L S_k^z \right)^2 + \frac{2}{G} \sum_{k=1}^L S_k^z -\frac{L^2}{2} I.$$
\subsection{First form of Bethe Ansatz solution}

The Bethe Ansatz solution for \eqref{eq:ppip0} was obtained in \cite{ilsz09}. We state the solution and its connection to the $\{t_j\}$ in \eqref{eq:ppip0tj1}: of the coupled Bethe Ansatz equations
\begin{equation} \label{eq:ppip0bae}
    \frac{G^{-1} + 2M - L - 1}{y_k} + \sum_{l=1}^L \frac{1}{y_k  - z_l^2} = \sum_{j \neq k}^M \frac{2}{y_k - y_j}, \qquad k=1,\dots ,M,
\end{equation}
where $M$ is the quantum number of particle-pairs, for each solution $\{ y_k \}$ known as the Bethe roots, there exists a correspondence
between $\{y_k\}$ and $\{t_j\}$ given by the following
\begin{align} \label{eq:ppip0rootstrans}
    t_j & = -G^{-1} - 2M + 2 z_j^2 \sum_{k=1}^M \frac{1}{z_j^2 - y_k}.
\end{align}
The techniques involved in achieving this connection \eqref{eq:ppip0rootstrans} first appeared in \cite{bt07} and are also adopted in \cite{fesg11, fs12, cdvv15, cvd17} later. The corresponding energy for \eqref{eq:ppip0} is given by
\begin{equation*}
E=\frac{G}{2} \sum_{j=1}^L z_j^2 t_j + \frac{1}{2}\sum_{j=1}^L z_j^2 = (1+G)\sum_{k=1}^M y_k.
\end{equation*}
Each eigenstate has the form
$$|\Phi \rangle = \prod_{k=1}^M C(y_k) |0\rangle, $$
where $|0\rangle$ denotes the vacuum state and 
$$C(y) = \sum_{j=1}^L \frac{z_j}{y-z_j^2} S_k^+. $$
\subsection{Second form of Bethe Ansatz solution}\label{sec:bae2}

Alternatively, a second form of Bethe Ansatz solution can be derived from the hole-pair perspective \cite{lmm15} for the isolated case. Let $P = L - M$ denote the quantum number of hole-pairs. For each solution $\{ v_k \}$ of the coupled equations
\begin{equation} \label{eq:ppip0bae2}
    \frac{-G^{-1} + 2P - L -1}{v_k} + \sum_{l=1}^L \frac{1}{v_k - z_l^2} = \sum_{j \neq k}^P \frac{2}{v_k - v_j}, \qquad k= 1, \dots,P,
\end{equation}
there exists a correspondence between $\{v_k\}$ and $\{t_j\}$ given by the following
\begin{align*}
    t_j & = G^{-1} - 2P + 2 z_j^2 \sum_{k=1}^P \frac{1}{z_j^2 - v_k}.
\end{align*}
The corresponding energy is given by
\begin{equation*}
    E =\frac{G}{2} \sum_{j=1}^L z_j^2 t_j + \frac{1}{2}\sum_{j=1}^L z_j^2 = \sum_{l=1}^L z_l^2 + (G-1) \sum_{k=1}^P v_k.
\end{equation*}
Each eigenstate has the form
\begin{equation*}
    |\Psi \rangle = \prod_{k=1}^P B(y_k) |\chi \rangle, 
\end{equation*}
where $|\chi \rangle$ denotes the completely filled state of $L$ particle-pairs and
\begin{equation*}
    B(y) = \sum_{j=1}^L \frac{z_j}{y-z_j^2} S_k^-. 
\end{equation*}
\subsection{Symmetries of Bethe Ansatz solutions} \label{sec:ppip0baesyms}

\begin{figure}
    \centering
\includegraphics[scale=0.9]{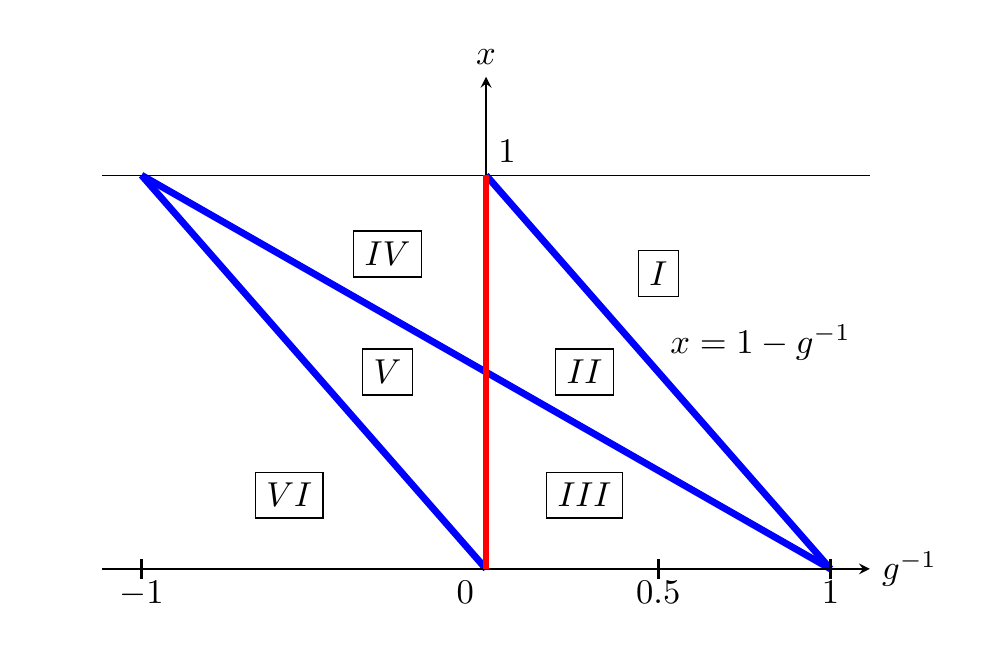}
\caption{Region $I$ with $x>1-g^{-1}$ is known as the weak coupling BCS phase. The boundary between $I$ and $II$, i.e. $x = 1-g^{-1}$, is known as the Moore-Read line. Region $II$ with $(1-g^{-1})/2< x< 1-g^{-1}$ and $g^{-1}>0$ is known as the weak pairing phase. The boundary between $II$ and $III$, i.e. $x = (1-g^{-1})/2$ is known as the Read-Green line. Region $III$ with $x<(1-g^{-1})/2$ and $g^{-1}>0$ is known as the strong pairing phase. The properties of the model for $g<0$ have attracted little attention.}
\label{fig:fig1}
\end{figure}

Introduce the following parameters,
\begin{align*}
    x & = M/L, \qquad g = GL,
\end{align*}
such that $x$ takes values in $[0,1]$ and $g$ in $(-\infty,\infty)$ as shown in Fig.\phantom{ }\ref{fig:fig1}.
There exists a rotational symmetry around the point $(g^{-1}=0, x=0.5)$ between the two forms of Bethe Ansatz solutions. For instance, if we have a solution $\{y_k\}$ to \eqref{eq:ppip0bae} with $x=0.6$, $g^{-1} = 0.4$, then this solution corresponds to a solution $\{v_k = y_k\}$ to \eqref{eq:ppip0bae2} with $x=0.4$,  $g^{-1} = -0.4$ and $\{z_l\}$ being fixed.

Apart from this correspondence, there exists another type of relation which we call inversion. Inversion establishes an invertible mapping, given by a skewed reflection against the line $x= 0.5 - g^{-1}$, between solution sets in regions $I$ and $VI$, and between solution sets in regions $II$ and $V$. Regions $IV$ and $III$ are stable under inversion. Consider the second form of Bethe Ansatz equations \eqref{eq:ppip0bae2}. By setting $v_k = u_k^{-1}$ we can derive the following expression,
\begin{equation*} 
    \frac{G^{-1} -1}{u_k} + \sum_{l=1}^L \frac{1}{u_k - z_l^{-2}} = \sum_{j \neq k}^P \frac{2}{u_k - u_j}, \qquad k= 1, \dots,P.
\end{equation*}
Note that under inversion, $\{z_l\}$ is no longer preserved. In other words, knowing a solution $\{ v_k \}$ to \eqref{eq:ppip0bae2} with a certain set of parameters $\{x,g^{-1},z_l\}$, we also obtain a solution $\{u_k= v_k^{-1}\}$ to \eqref{eq:ppip0bae2} with transformed parameters $\{x, \bar{g}^{-1} = -2x-g^{-1}+1, \bar{z}_l = z_l^{-1}\}$.

When we combine the rotational symmetry and inversion, we achieve a correspondence between solution sets to \eqref{eq:ppip0bae} and \eqref{eq:ppip0bae2}. For instance, a solution $\{y_k\}$ to \eqref{eq:ppip0bae} with $x=0.6$, $g^{-1}=0.4$ and parameters $\{z_l\}$ is also a solution $\{v_k=y_k\}$ to \eqref{eq:ppip0bae2} with $x=0.4$, $g^{-1}= -0.4$ and the same parameters $\{z_l\}$. Then the inversion of this solution $\{u_k=v_k^{-1}\}$ also solves \eqref{eq:ppip0bae2} with $x=0.4$, $g^{-1}=0.6$ and transformed parameters $\{z_l^{-1}\}$. Moreover, since $G$, $\Gamma$ and $z_k$ are all positive real constants, the Bethe Ansatz equations \eqref{eq:ppip0bae} and \eqref{eq:ppip0bae2} are invariant under complex conjugation. This implies that, in the absence of degeneracy in the spectrum of the set of conserved operators, every solution set of Bethe roots consists of complex-conjugate pairs or real numbers. 
\subsection{Integral approximation for the first form of Bethe Ansatz solution} \label{sec:bae1cla}

Numerical solution for the ground-state Bethe roots $\{y_j\}$ in \eqref{eq:ppip0bae} and its peculiar behaviour under certain choices of parameters are discussed in \cite{ilsz09} and \cite{vdv14}.  The distribution of the Bethe roots suggests that they lie on curves in the complex plane, allowing an integral approximation to be applied. The continuum limit approximation for \eqref{eq:ppip0bae} where $L$ is large is studied in \cite{dilsz10,rdo10}. In the limit, we require that $M$ also becomes large while $x$ is finite, and similarly $G$ becomes small while $g$ is finite.

First we formally define the discrete density for each single particle energy level $z_k^2$. Since all the $z_k^2$ are real and positive it is natural to relabel them as $z_k^2$ with $z_k^2 < z_j^2$ whenever $k<j$ such that $z_1^2$ is the smallest and $z_L^2$ is the largest. The discrete root density $\tilde{\rho}$ is defined as
\begin{equation} \label{eq:cladef}
    \tilde{\rho}(z_j^2) = \frac{L}{(L-1)(z_{j+1}^2 - z_j^2)}, \qquad \tilde{\rho}(z_{L}^2) = 0,
\end{equation}
such that
\begin{equation*}
    \sum_{j=1}^L \tilde{\rho}(z_j^2)\cdot (z_{j+1}^2 - z_j^2) = L.
\end{equation*}
In the continuum limit, we introduce $\rho$ to be the continuum density for $z_k^2$ with connected support being a subset of $(0,\omega)$ and replace all $z_k^2$ with a continuous variable $\epsilon$.  the continuum approximation for summation over any given function $f$ is undertaken by replacing sum with integral according to
\begin{equation} \label{eq:cladef2}
    \sum_{j=1}^L f(z_j^2) \longrightarrow \int_0^\omega {\rm d} \epsilon \, \rho(\epsilon) f(\epsilon).
\end{equation}
Setting $f=1$ in \eqref{eq:cladef2} gives the normalization condition for the density,
\begin{equation} \label{eq:ppip0cladensity}
    \int_0^\omega {\rm d} \epsilon \, \rho(\epsilon) = L.
\end{equation}
We also introduce a continuous curve $\Omega$, which is invariant under complex conjugation, to approximate the distribution of the ground-state Bethe roots $\{y_k\}$ in \eqref{eq:ppip0bae}. Let $r(y)$ be the density for $\{y_k\}$ in the continuum limit with support on $\Omega$. Since on the RHS of \eqref{eq:ppip0bae}, the expression
$$f(y_j) =  \frac{2}{y_k - y_j} $$
gives rise to a singularity, we adopt the Cauchy principal value to approximate the summation
\begin{equation*}
    \sum_{j \neq k} ^ M f(y_j) \longrightarrow \dashint_\Omega |{\rm d}y| \, r(y) f(y),
\end{equation*}
where $\dashint$ denotes the Cauchy principal value of an integral. The continuum limit approximation for the Bethe Ansatz solution $\{y_k\}$ for \eqref{eq:ppip0bae} reads as
\begin{align} \label{eq:ppip0baecla}
\int_0^\omega {\rm d} \epsilon \, \frac{\rho(\epsilon)}{\epsilon -y} - \frac{G^{-1}+2M-L}{y} & = \dashint_\Omega |{\rm d}y'| \, \frac{2r(y')}{y'-y},\\
\label{eq:ppip0baecla2}
\int_\Omega |{\rm d}y| \, r(y) & = M,
\end{align}
where equation \eqref{eq:ppip0baecla2} is the normalization condition for the density $r(y)$. The ground-state energy is given by
$$ E = \int_\Omega |{\rm d} y| \, y r(y). $$

\begin{figure}
    \centering
 \includegraphics[scale=1]{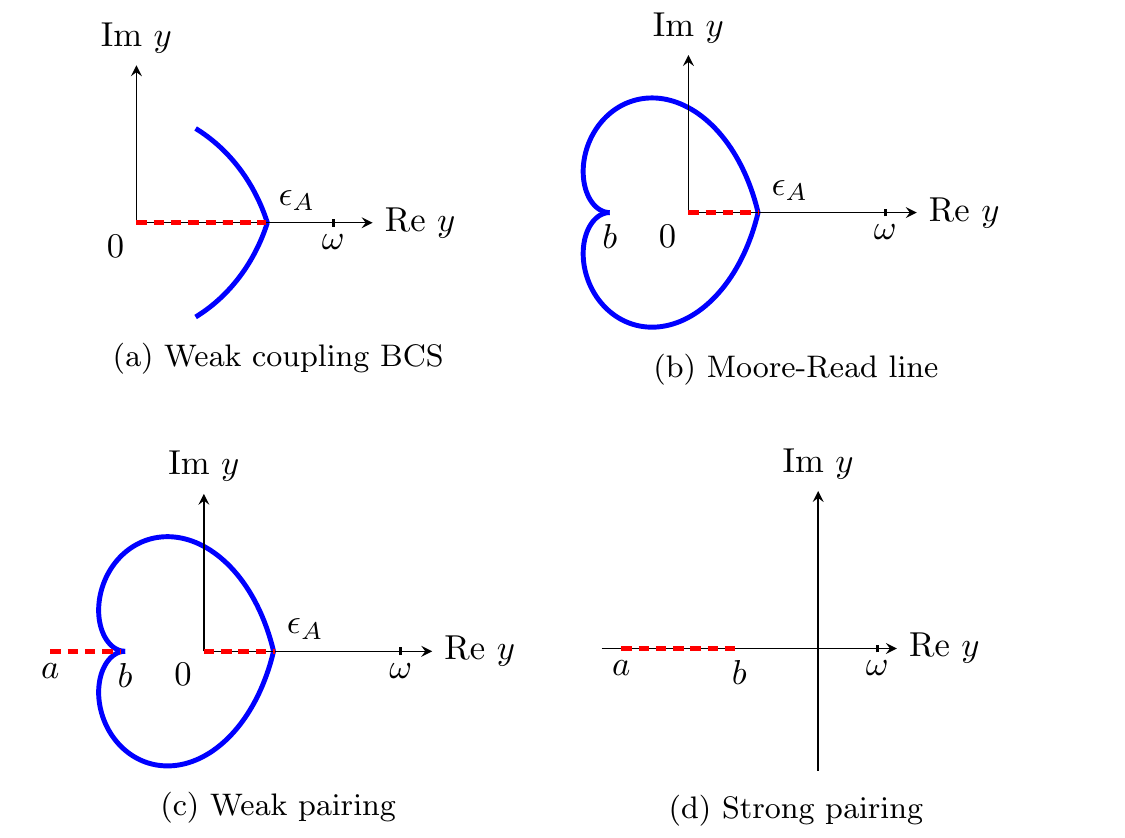}    
\caption{The solution curve $\Omega$ evolves from (a) to (d) as we send $g$ from $0$ to $+\infty$ while $x$ is fixed at a value between $0$ and $0.5$.}
\label{fig:fig2}
\end{figure}

The solution curve $\Omega$ and $r(y)$ have been solved in \cite{dilsz10} and are dependent on the choice of the parameters $x$, $g^{-1}$ as shown in Fig.\phantom{ }\ref{fig:fig1}. The solution curve $\Omega$ consists of two parts, a complex part $\Omega_C$ depicted by a solid line and a real part $\Omega_A$ depicted by one or more dashed lines, see Fig.\phantom{ }\ref{fig:fig2}. In Fig.\phantom{ }\ref{fig:fig2}(a), the complex arc $\Omega_C$ intersects the real part $\Omega_A=(0,\epsilon_A)$ at $\epsilon_A$.  In Fig.\phantom{ }\ref{fig:fig2}(b), the arc $\Omega_C$ closes with an endpoint $b$ on the negative real line as $(g^{-1},x)$ arrives at the Moore-Read line from right. In Fig.\phantom{ }\ref{fig:fig2}(c), a line segment $(a,b)$ forms on the negative real line adding a component to $\Omega_A$. Then the point $b$ will approach $0$ with the complex curve $\Omega_C$ shrinking until it vanishes when $b= \epsilon_A =0$ and $(g^{-1},x)$ arrives at the Read-Green line from right. In Fig.\phantom{ }\ref{fig:fig2}(d), as $(g^{-1},x)$ departs from the Read-Green line to the left, $b$ becomes negative and $\Omega$ consists of one real part $\Omega_A=(a,b)$.
\subsection{Approximation for the second form of Bethe Ansatz solution} \label{sec:bae2cla}

Following the approach as discussed in previous Sect.\phantom{ }\ref{sec:bae1cla}, the continuum limit approximation for the ground-state Bethe Ansatz solution $\{v_k\}$ of \eqref{eq:ppip0bae2} reads as
\begin{align} \label{eq:ppip0bae2cla}
\int_0^\omega {\rm d}\epsilon \, \frac{\rho(\epsilon)}{\epsilon - v} - \frac{-G^{-1} + 2P-L}{v} & = \dashint_{\Omega} |{\rm d}v'| \, \frac{2r(v')}{v'-v},
\end{align}
where $\Omega$ is a continuous curve introduced to approximate the distribution of the Bethe roots $\{v_k\}$, and $r(v)$ is the density for $\{v_k\}$ in the continuum limit satisfying
$$
\int_{\Omega} |{\rm d}v| \, r(v) = P.
$$
The ground-state energy is given by
\begin{align*}
E & = \int_0^\omega {\rm d} \epsilon \, \epsilon \rho(\epsilon) - \int_{\Omega} |{\rm d}v| \, vr(v).
\end{align*}

\begin{figure}
    \centering
\includegraphics[scale=0.95]{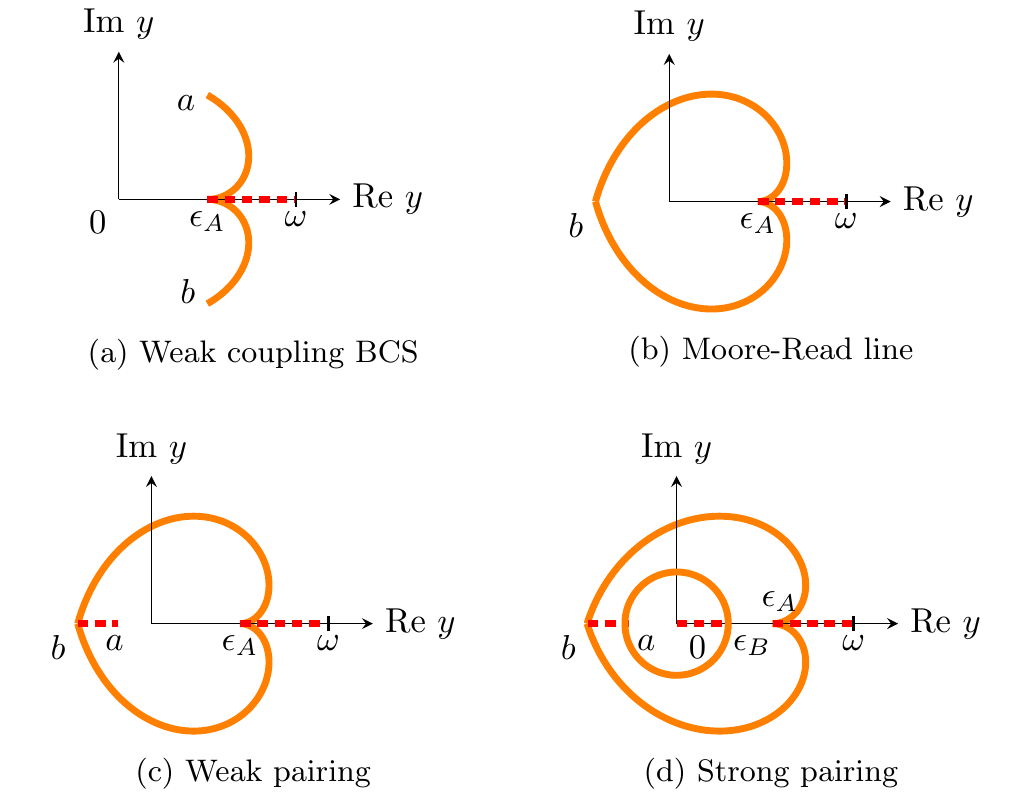}    
\caption{The solution curve $\Omega$ proposed for the integral approximation for the second form Bethe Ansatz solution consists of a real part $\Omega_A$ depicted by one or more dashed lines and a complex part $\Omega_C$ depicted by one or more solid lines. Again we fix $x \in (0,0.5)$ and send $g$ from $0$ to $+\infty$.}
\label{fig:fig3}
\end{figure}

Now we need to solve for $\Omega$ and density $r(v)$ for the ground state. The solution curve $\Omega$ is proposed to be classified under the four phases as of the approximation for the first form Bethe Ansatz solution \eqref{eq:ppip0baecla}, see Fig.\phantom{ }\ref{fig:fig3}.
Here we have modified shapes deduced from the rotational symmetry combined with inversion discussed in Sect.\phantom{ }\ref{sec:ppip0baesyms}. Fig.\phantom{ }\ref{fig:fig3}(a)(b)(c) are topological inversion of Fig.\phantom{ }\ref{fig:fig2}(a)(b)(c). In Fig.\phantom{ }\ref{fig:fig3}(c), the point $a$ approaches zero as $(g^{-1},x)$ approaches the Read-Green line from right. As $(g^{-1},x)$ departs from the Read-Green line and continues to move left, a third real line segment $(0,\epsilon_B)$ and a second complex loop intersecting the real line at $a$ and $\epsilon_B$ appear, see Fig.\phantom{ }\ref{fig:fig3}(d). Further calculation gives rise to the same energy expression as from the integral approximation \eqref{eq:ppip0baecla} and mean-field analysis \cite{dilsz10}.
\subsection{Limitations of the continuum limit approximation for the Bethe Ansatz solutions} \label{sec:baeclalmt}

The Moore-Read line is an example of a ground-state phase boundary line associated with changes in the topology of the root distribution. In previous Sect.\phantom{ }\ref{sec:bae1cla} and \ref{sec:bae2cla}, as we send parameters $(g^{-1},x)$ from the weak coupling BCS phase to the Moore-Read line, the complex part of the solution curve $\Omega_C$ evolves until it closes and forms a loop. However in the discrete case, all the Bethe roots condense at the origin \cite{ilsz09} when the parameters reach the Moore-Read line. This discrepancy between the integral approximation and the discrete case is not present in models such as the Richardson $s$-wave pairing model and the $d+id$-wave pairing Hamiltonian \cite{ml12,ml13} where a similar approach of integral approximation is adopted. In the case of the two-level Richardson $s$-wave pairing model \cite{ml12}, the solution curve of the integral approximation evolves until it closes and forms a loop as some governing parameters approach a ground-state phase boundary line, meanwhile the discrete Bethe roots do not condense and their distribution is predicted by the solution curve within small error in the limit. In the case of the $d+id$-wave pairing Hamiltonian \cite{ml12,ml13}, the solution curve of the integral approximation contracts to a point at the origin and the discrete Bethe roots also condense at the origin as the governing parameters approach a ground-state phase boundary line.

Due to the aforementioned discrepancy in the closed model, we perform a closer inspection of the integral approximation \eqref{eq:ppip0baecla} in the Moore-Read line case with solution curve $\Omega$ depicted in Fig.\phantom{ }\ref{fig:fig2}(b). The general form of the density $r(y)$ for this case is proposed in \cite{dilsz10} to be the following,
\begin{align} \label{eq:ppip0baeclarsrho}
r(y) \, |{\rm d}y| & =
  \begin{cases}
    s(y) \, {\rm d}y     & \quad \text{if } y \in \Omega_C\\
    \rho(y) \, {\rm d}y  & \quad \text{if } y \in \Omega_A
  \end{cases},\\ \nonumber
    s(y) & = \frac{1}{2i\pi} \left[ \int_0^\omega {\rm d} \epsilon \, \frac{\rho(\epsilon)}{\epsilon - y} + \frac{M}{y} \right],\\ \nonumber
    s(a) & = 0.
\end{align}
The arc $\Omega_C$ is obtained by solving the following integral equation
\begin{equation} \label{eq:ppip0gammac}
    {\rm Im} \left[ \int_{y_0}^y {\rm d}y' \,  s(y') \right] = 0 , \quad y \in \Omega_C,
\end{equation}
where $y_0$ is any point of $\Omega_C$. 

We first consider the limiting case where $G \rightarrow 0$. In this case the Bethe roots $\{y_k\}$ are all real and lie within the interval $(0,\omega_f]$ where $\omega_f$ is the upper bound for $\{ y_k \}$ and $ \omega_f < \omega$. Hence in the integral approximation the solution curve $\Omega$ reduces to $\Omega_A = (0,\omega_f)$ and $\epsilon_A = \omega_f$. According to \eqref{eq:ppip0baeclarsrho}, the normalization condition \eqref{eq:ppip0baecla2} then reads as
\begin{equation} \label{eq:ppip0baecla3}
    \int_\Omega |{\rm d}y| \, r(y) = \int_0^{\omega_f} {\rm d}y \, \rho(y) = M,
\end{equation}
which determines the value of $\omega_f$. As $G$ increases, the complex part $\Omega_C$ starts to form and $\epsilon_A$ decreases away from $\omega_f$. Consequently the allowable bound for $\epsilon_A$ is $(0,\omega_f)$.

\begin{figure}[ht]
    \begin{subfigure}{0.49\textwidth}
    \includegraphics[scale=0.23]{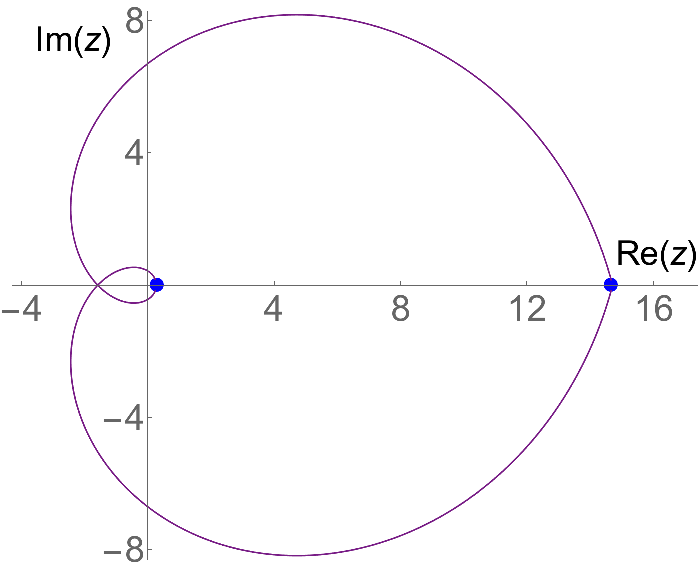}
    \caption{}
    \end{subfigure}
    \begin{subfigure}{0.49\textwidth}
    \includegraphics[scale=0.23]{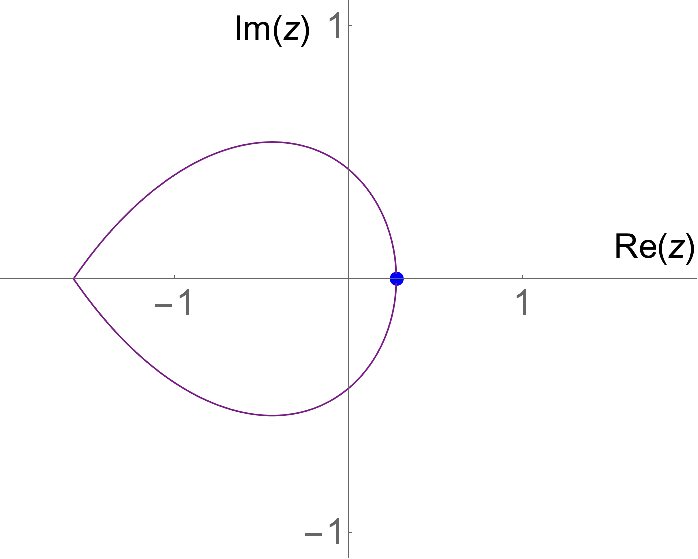}
    \caption{}
    \end{subfigure}
\caption{In (a), equation \eqref{eq:ppip0gammac1} is plotted with $z=y/\omega_0$ and the filling fraction $x=0.4$. The inner loop and the outer loop are two candidates for $\Omega_C$. Note that in our setting $\omega_0 < \epsilon_A < \omega_{f} = 5\omega_0/3$, hence $1< \epsilon_A/\omega_0 < 5/3
$. Numerical calculation shows that $\epsilon_A/\omega_0 \approx 14.71$ or $0.2748$, depicted by solid dots. In (b), the inner loop is plotted with an enlarged scale.}
\label{fig:fig4}
\end{figure}

Now we consider the special case of an inverse-square density
\begin{align*}
    \rho(\epsilon) & = \frac{\omega_0 L}{\epsilon^2}, \quad \epsilon \in (\omega_0, +\infty).
\end{align*}
This is a limiting case for the density $\rho(\epsilon)$ as we let it vanish on $(0,\omega_0)$ while sending $\omega \rightarrow +\infty$. When $G \rightarrow 0$, substituting the inverse-square density into \eqref{eq:ppip0baecla3} we have $\omega_f = \omega_0/(1-x)$. Hence the constraint for $\epsilon_A$ is
\begin{equation} \label{eq:allowablebound}
    \epsilon_A \in \left( \omega_0, \frac{\omega_0}{1-x} \right).
\end{equation}
The equation for $\Omega_C$ is then derived from \eqref{eq:ppip0gammac},
\begin{align*}
    {\rm Re}\left[ \omega_0 L\int_{\omega_0}^\infty {\rm d} \epsilon \, \frac{1}{\epsilon^2} \log{\left( \frac{\epsilon - y}{\epsilon - y_0}\right)} - M \log{\frac{y}{y_0}} \right] = 0.
\end{align*}
Since
\begin{multline*}
    \omega_0 \int_{\omega_0}^\infty {\rm d} \epsilon \, \frac{1}{\epsilon^2} \log{\left(\frac{\epsilon-y}{\epsilon-y_0} \right)}\\
    \qquad = \left( 1-\frac{\omega_0}{y} \right) \log{\left( 1 - \frac{y}{\omega_0} \right)} - \left( 1 - \frac{\omega_0}{y_0} \right) \log{\left( 1 - \frac{y_0}{\omega_0} \right)},
\end{multline*}
the arc $\Omega_C$ is determined by the following equation
\begin{equation*}
    {\rm Re} \left[ \left( 1-\frac{\omega_0}{y} \right) \log{\left( 1 - \frac{y}{\omega_0} \right)} - \left( 1 - \frac{\omega_0}{y_0} \right) \log{\left( 1 - \frac{y_0}{\omega_0} \right)} - x \log{\frac{y}{y_0}} \right] = 0.
\end{equation*}
Choosing $y_0$ to be $a$, $\Omega_C$ is given by
\begin{equation} \label{eq:ppip0gammac1}
    {\rm Re} \left[ \left( 1-\frac{\omega_0}{y} \right) \log{\left( 1 - \frac{y}{\omega_0} \right)} - \left( 1 - \frac{\omega_0}{a} \right) \log{\left( 1 - \frac{a}{\omega_0} \right)} - x \log{\frac{y}{a}} \right] = 0.
\end{equation}
The remaining constraint $s(a) = 0$ from \eqref{eq:ppip0baeclarsrho} implies that
\begin{align*}
    \omega_0 L\int_{\omega_0}^\infty \frac{1}{(\epsilon-a)\epsilon^2} + \frac{M}{a} & = 0,
\end{align*}
hence we have
\begin{align} 
    \label{eq:ppip0gammac2}
    \exp{\left((1-x)\frac{-a}{\omega_0}\right)} & = 1 + \frac{-a}{\omega_0}.
\end{align}
Setting $x = 0.4$, from \eqref{eq:ppip0gammac2} we numerically determine that $-a/\omega_0 \approx 1.58$. From \eqref{eq:ppip0gammac1}, the solution curve for $y/\omega_0$ is plotted as in Fig.\phantom{ }\eqref{fig:fig4}.

However, with this choice of parameters and inverse-square density, numerical results show that the intersection $\epsilon_A$ is outside the allowable bound $(\omega_0,5\omega_0/3)$ from \eqref{eq:allowablebound}. This inconsistency is verified when we study the solution curve in the integral approximation for the second form of Bethe Ansatz solution \eqref{eq:ppip0bae2cla} in the Moore-Read line case with a constant density, which corresponds to an inverse-square density for \eqref{eq:ppip0baecla} in the Moore-Read line case as is discussed in Sect. \ref{sec:ppip0baesyms}. The resulting solution curve again intersects the real line at positions outside the allowable bound.

Alternatively, if we choose a constant density $\rho(\epsilon)=L/(\omega-\omega_0)$ with support on $(\omega_0, \omega)$ for \eqref{eq:ppip0baecla}, it can be shown that for the integral approximation to be valid $\omega_0/\omega$ cannot exceed the numerically determined value $0.04246$. These results suggest that the validity of the continuum limit approximation is dependent on the choice of the density function $\rho(\epsilon)$.
\section{Conserved operator eigenvalue method} \label{sec:coem}

The difficulties in the integral approximation for the Bethe roots $\{y_k\}$ arise from the task to find a suitable solution curve $\Omega$ and solve for its density $r(y)$ in \eqref{eq:ppip0baecla}.
The next step is to adopt an alternative approach that eliminates these requirements and accommodates to an arbitrary form of density $\rho(\epsilon)$ subject to \eqref{eq:ppip0cladensity}. In the following discussion, we continue with the approach that first appeared in \cite{bt07}. Defining 
$$ \Lambda_j = \sum_{k=1}^M \frac{1}{z_j^2 - y_k},$$
from \eqref{eq:ppip0rootstrans} we have
\begin{align} \label{eq:ppip0rootstrans2}
    t_j & = -G^{-1} - 2M + 2z_j^2 \Lambda_j.
\end{align}
Since each solution $\{y_k\}$ to \eqref{eq:ppip0bae} consists of complex-conjugate pairs or real numbers, $\Lambda_j$ and $t_j$ are all real. Substituting into \eqref{eq:ppip0tj1} we obtain the following quadratic equations \cite{cdvv15, cvd17},
\begin{equation} \label{eq:ppip0baelambda}
    \Lambda_j^2 -\frac{2q}{z_j^2} \Lambda_j = \sum_{k\neq j}^L \frac{\Lambda_j - \Lambda_k}{z_j^2 - z_k^2} + \frac{1}{z_j^2} \sum_{l=1}^{L} \Lambda_l,
\end{equation}
where $2q =  G^{-1} + 2M - L - 1$. A continuum limit approximation applied to \eqref{eq:ppip0baelambda} leads to the following integral equation,
\begin{equation} \label{eq:ppip0prop1eq}
\Lambda (\epsilon)^2 - \frac{2q}{\epsilon} \Lambda (\epsilon) = \int_0^\omega {\rm d} \delta \, \rho (\delta) \frac{\Lambda (\epsilon) -\Lambda (\delta) }{\epsilon - \delta}  + \frac{1}{\epsilon} \int_0^\omega {\rm d} \delta \, \rho(\delta) \Lambda(\delta).
\end{equation}
The objective now is to derive a solution to \eqref{eq:ppip0prop1eq}. We will establish several useful results before the solution is stated. Let
\begin{equation} \label{eq:ppip0claconstraints}
    R(\epsilon) = \sqrt{(\epsilon - a )( \epsilon - b)}, \quad \int_0^\omega {\rm d} \epsilon \, \frac{\rho(\epsilon)}{R(\epsilon)} = \frac{2|q|}{\sqrt{ab}}, \quad \int_0^\omega {\rm d} \epsilon \, \frac{\epsilon \rho(\epsilon)}{R(\epsilon)} =  \frac{1}{G}. 
\end{equation}
Note that $a, b$ are determined by $G$ and $q$. We require them to be a complex-conjugate pair, or both negative real numbers, i.e.
\begin{equation} \label{eq:ppip0claconstraints2}
 a = \bar{b} \in \mathbb{C} \setminus \mathbb{R} \quad \text{or} \quad a \leq b \leq 0. 
\end{equation}
As a result $a+b\in \mathbb{R}$ and $ab \geq 0$. The function $R(\epsilon)$ is the elementary real-valued square-root with $\epsilon \in (0,\omega)$. In the limit where $q = 0$, we require $b$ to vanish also. In this case \eqref{eq:ppip0claconstraints} becomes
\begin{equation} \label{eq:ppip0claconstraints3}
    R(\epsilon) = \sqrt{\epsilon ( \epsilon - a)}, \quad \int_0^\omega {\rm d}\epsilon \, \frac{\rho(\epsilon)}{R(\epsilon)} = C, \quad \int_0^\omega {\rm d} \epsilon \, \frac{\epsilon \rho(\epsilon)}{R(\epsilon)} =  \frac{1}{G},
\end{equation}
where $a$ is real and determined by $G$. All the following results and proofs leading to the expression for the ground-state energy for the closed model assume $q \neq 0 \neq b$. In the special case where $q=b=0$, simply replace $(2|q|/\sqrt{ab})$ with the finite value $C$ in \eqref{eq:ppip0claconstraints3}, the proof of which follows a similar calculation and is omitted.
\begin{lemma}
For all $\gamma,\delta,\epsilon \in (0,\omega)$, the following identity holds
\begin{equation} \label{eq:lemma1}
 \frac{R(\gamma)-R(\epsilon)}{(\gamma-\epsilon)(\delta -\epsilon)} - \frac{R(\gamma)-R(\delta)}{(\gamma-\delta)(\delta -\epsilon)} = \frac{R(\delta)-R(\epsilon)}{(\delta-\epsilon)(\gamma - \epsilon)} - \frac{R(\delta)-R(\gamma)}{(\delta - \gamma)(\gamma - \epsilon)}.
\end{equation}
Proof: Let
$$ \theta_L(\gamma,\delta,\epsilon) =  \frac{R(\gamma)-R(\epsilon)}{(\gamma-\epsilon)(\delta -\epsilon)} - \frac{R(\gamma)-R(\delta)}{(\gamma-\delta)(\delta -\epsilon)}, $$
$$ \theta_R(\gamma,\delta,\epsilon) = \frac{R(\delta)-R(\epsilon)}{(\delta-\epsilon)(\gamma - \epsilon)} - \frac{R(\delta)-R(\gamma)}{(\delta - \gamma)(\gamma - \epsilon)}. $$
For $\delta \neq \gamma \neq \epsilon \neq \delta$,  $$ \theta_L(\gamma,\delta,\epsilon) - \theta_R(\gamma,\delta,\epsilon) = 0, $$ the proof of which is straightforward and omitted here. It can be shown that
\begin{multline*}
    \theta_L (\gamma,\delta,\epsilon)  = \frac{1}{\big(R(\gamma) + R(\epsilon)\big) \big(R(\gamma)+R(\delta)\big)} \\
     \quad \times \bigg( -R(\gamma) + \big(\gamma - (a+b)\big) \frac{\delta + \epsilon - (a+b)}{R(\delta) + R(\epsilon)} + \frac{(a+b)\epsilon\delta - ab(\delta + \epsilon)}{\epsilon R(\delta) + \delta R(\epsilon)}  \bigg).
\end{multline*}
This representation of $\theta_L(\gamma,\delta,\epsilon)$ has no singularities for all $\gamma,\delta,\epsilon \in (0,\omega)$. The same expression can be found for 
$\theta_R(\gamma,\delta,\epsilon)$. Therefore $\theta_L(\gamma,\delta,\epsilon) - \theta_R(\gamma,\delta,\epsilon)$ continuously extends to vanish for all $\gamma,\delta,\epsilon \in (0,\omega)$.\\
\qed
\end{lemma}
\begin{corollary} \label{cor1}
Let
\begin{align*}
    \mathcal{I} & = 2\int_0^\omega \int_0^\omega  {\rm d} \delta \, {\rm d} \gamma \,  \frac{\rho(\gamma)\rho(\delta)}{R(\delta)}\left( \frac{R(\delta) - R(\epsilon)}{(\delta - \epsilon)(\gamma - \epsilon)} - \frac{R(\delta) - R(\gamma)}{(\delta - \gamma)(\gamma - \epsilon)} \right), \\
    \mathcal{J} & = \left( \int_0^\omega {\rm d} \gamma \, \frac{\rho (\gamma)}{R(\gamma)} \frac{R(\gamma)-R(\epsilon)}{\gamma-\epsilon} \right)^2,
\end{align*}
then 
$$ \mathcal{I} = \mathcal{J} - \frac{4q^2}{ab}.$$
Proof: Exchanging the  variables $\delta$ and $\gamma$ in the integrand, and then performing a change of order of integration yields
\begin{align*}
\mathcal{I} & = 2\int_0^\omega \int_0^\omega {\rm d} \delta \, {\rm d} \gamma \, \frac{\rho(\gamma)\rho(\delta)}{R(\gamma)}\left( \frac{R(\gamma) - R(\epsilon)}{(\gamma - \epsilon)(\delta - \epsilon)} - \frac{R(\gamma) - R(\delta)}{(\gamma - \delta)(\delta - \epsilon)} \right) \\
& = 2\int_0^\omega \int_0^\omega {\rm d} \delta \, {\rm d} \gamma \, \frac{\rho(\gamma)\rho(\delta)}{R(\gamma)}\left( \frac{R(\delta) - R(\epsilon)}{(\delta - \epsilon)(\gamma - \epsilon)} - \frac{R(\delta) - R(\gamma)}{(\delta - \gamma)(\gamma - \epsilon)} \right),
\end{align*}
where the second step is due to \textbf{Lemma 1}. Now we add up two distinct expressions for $\mathcal{I}/2$,
\begin{align*}
    \mathcal{I} & = \int_0^\omega \int_0^\omega  {\rm d} \delta \, {\rm d} \gamma \,  \frac{\rho(\gamma)\rho(\delta)}{R(\delta)}\left( \frac{R(\delta) - R(\epsilon)}{(\delta - \epsilon)(\gamma - \epsilon)} - \frac{R(\delta) - R(\gamma)}{(\delta - \gamma)(\gamma - \epsilon)} \right)\\
    & \quad + \int_0^\omega \int_0^\omega {\rm d} \delta \, {\rm d} \gamma \, \frac{\rho(\gamma) \rho(\delta)}{R(\gamma)}\left( \frac{R(\delta) - R(\epsilon)}{(\delta - \epsilon)(\gamma - \epsilon)} - \frac{R(\delta) - R(\gamma)}{(\delta - \gamma)(\gamma - \epsilon)} \right)\\
    & = \int_0^\omega \int_0^\omega  {\rm d} \delta \, {\rm d} \gamma \, \rho(\gamma)\rho(\delta) \frac{R(\delta) + R(\gamma)}{R(\delta)R(\gamma)}\left( \frac{R(\delta) - R(\epsilon)}{(\delta - \epsilon)(\gamma - \epsilon)} - \frac{R(\delta) - R(\gamma)}{(\delta - \gamma)(\gamma - \epsilon)} \right).
\end{align*}
Since
\begin{multline*}
    \big(R(\delta) + R(\gamma)\big) \left( \frac{R(\delta)-R(\epsilon)}{(\delta - \epsilon)(\gamma - \epsilon)} - \frac{R(\delta) - R(\gamma)}{(\delta - \gamma)(\gamma - \epsilon)} \right)\\
    = -1 + \frac{\big(R(\gamma) - R(\epsilon)\big)\big(R(\delta) - R(\epsilon)\big)}{(\gamma - \epsilon)(\delta - \epsilon)},
\end{multline*}
we conclude that
\begin{align*}
    \mathcal{I} & = \int_0^\omega {\rm d} \delta \, \frac{\rho(\delta)}{R(\delta)} \frac{R(\delta) - R(\epsilon)}{\delta - \epsilon}  \int_0^\omega {\rm d} \gamma \, \frac{\rho(\gamma)}{R(\gamma)} \frac{R(\gamma) - R(\epsilon)}{\gamma - \epsilon}\\
    & \quad - \int_0^\omega \int_0^\omega  {\rm d} \delta \, {\rm d} \gamma \, \frac{\rho(\delta)\rho(\gamma)}{R(\delta) R(\gamma)}\\
    & = \mathcal{J} - \frac{4q^2}{ab}.
\end{align*}
\qed
\end{corollary}
\begin{lemma} \label{lemma1}
Let
\begin{align*}
    \mathcal{K} & = \frac{2}{\epsilon} \int_0^\omega \int_0^\omega {\rm d}\gamma \, {\rm d}\delta \, \frac{\rho(\gamma)\rho(\delta)}{R(\gamma)} \frac{R(\gamma) - R(\delta)}{\gamma - \delta},\\
    \mathcal{O} & = \int_0^\omega {\rm d}\delta \, \rho(\delta)  \frac{R(\delta) - R(\epsilon)}{(\delta - \epsilon)\epsilon}  + \int_0^\omega  {\rm d}\delta \, \frac{\rho(\delta)R(\epsilon)}{R(\delta)} \frac{R(\delta) - R(\epsilon)}{(\delta - \epsilon)\epsilon},\\
    \mathcal{P} & = \int_0^\omega \int_0^\omega {\rm d}\epsilon \, {\rm d}\delta \, \rho(\epsilon) \rho(\delta) \frac{\epsilon \delta}{R(\delta)} \frac{R(\epsilon) - R(\delta)}{\epsilon - \delta},
\end{align*}
then 
\begin{align*}
    \mathcal{K} & = \frac{4|q|}{G\sqrt{ab}} \frac{1}{\epsilon} - \frac{4q^2}{ab} \frac{a+b}{\epsilon}, \\
    \mathcal{O} & = \frac{1}{G \epsilon} - \frac{2|q|}{\sqrt{ab}} \frac{a+b}{\epsilon} + \frac{2|q|}{\sqrt{ab}}, \\
    \mathcal{P} & = \frac{a + b}{2G^2} - \frac{2|q| \sqrt{ab}}{G} + \frac{1}{G} \int_0^\omega {\rm d}\epsilon \, \rho(\epsilon) R(\epsilon).
\end{align*}
\noindent Proof: Since
\begin{align*}
    \mathcal{K} & = \frac{2}{\epsilon} \int_0^\omega \int_0^\omega {\rm d}\gamma \, {\rm d}\delta \, \frac{\rho(\gamma)\rho(\delta)}{R(\gamma)} \frac{R(\gamma) - R(\delta)}{\gamma - \delta}\\
    & = \frac{2}{\epsilon} \int_0^\omega \int_0^\omega {\rm d}\gamma \, {\rm d}\delta \, \frac{\rho(\delta)\rho(\gamma)}{R(\delta)} \frac{R(\delta) - R(\gamma)}{\delta - \gamma},
\end{align*}
again by adding up two distinct expressions for $\mathcal{K}/2$, we have
\begin{align*}
    \mathcal{K} & = \frac{1}{\epsilon} \int_0^\omega \int_0^\omega {\rm d}\gamma \, {\rm d}\delta \, \frac{\rho(\gamma)\rho(\delta)}{R(\gamma)R(\delta)}\Big( R(\gamma) + R(\delta) \Big) \frac{R(\gamma) - R(\delta)}{\gamma - \delta} \\
    & = \frac{4|q|}{G\sqrt{ab}} \frac{1}{\epsilon} - \frac{4q^2}{ab} \frac{a+b}{\epsilon}.
\end{align*}
Next
\begin{align*}
\mathcal{O} & = \int_0^\omega {\rm d}\delta \, \frac{\rho(\delta)}{R(\delta)}\frac{1}{(\delta - \epsilon) \epsilon }\Big( R(\delta)^2 - R(\delta)R(\epsilon) +R(\epsilon)R(\delta) - R(\epsilon)^2 \Big) \\
& = \frac{1}{G\epsilon} + \frac{2|q|}{\sqrt{ab}} - \frac{2|q|}{\sqrt{ab}} \frac{a+b}{\epsilon}.
\end{align*}
Finally, following a similar calculation as that for $\mathcal{K}$,
\begin{align*}
    \mathcal{P} & = \frac{1}{2} \int_0^\omega \int_0^\omega {\rm d}\epsilon \, {\rm d}\delta \, \rho(\epsilon) \rho(\delta) \epsilon \delta \left( \frac{1}{R(\delta)} + \frac{1}{R(\epsilon)} \right) \frac{R(\epsilon) - R(\delta)}{\epsilon - \delta}\\
    & = \frac{a + b}{2G^2} - \frac{2|q| \sqrt{ab}}{G} + \frac{1}{G} \int_0^\omega {\rm d}\epsilon \, \rho(\epsilon) R(\epsilon) .
\end{align*}
\qed
\end{lemma}
Now we prove the following:
\begin{prop} \label{prop1}
The following function
\begin{equation} \label{eq:ppip0prop1sol}
\Lambda (\epsilon) = \frac{|q|}{\sqrt{ab}} \frac{R(\epsilon)}{\epsilon} + \frac{q}{\epsilon} - \frac{1}{2} \int_0^\omega {\rm d} \gamma \, \frac{\rho (\gamma)}{R(\gamma)} \cdot \frac{R(\gamma) - R(\epsilon)}{\gamma - \epsilon}
\end{equation}
is a solution to the integral equation \eqref{eq:ppip0prop1eq} with $a,b$ subject to \eqref{eq:ppip0claconstraints} and \eqref{eq:ppip0claconstraints2}.\\
Proof: \begin{align*}
\Lambda(\epsilon)^2 = \frac{\mathcal{J}}{4} & + \frac{q^2}{ab} \frac{R(\epsilon)^2}{\epsilon^2} + \frac{q^2}{\epsilon^2} - \frac{|q|}{\sqrt{ab}} \frac{R(\epsilon)}{\epsilon} \int_0^\omega {\rm d}\gamma \, \frac{\rho(\gamma)}{R(\gamma)} \frac{R(\gamma)-R(z)}{\gamma-z} \\
& - \frac{q}{\epsilon} \int_0^\omega {\rm d}\gamma \, \frac{\rho(\gamma)}{R(\gamma)} \frac{R(\gamma) - R(\epsilon)}{\gamma - \epsilon} +\frac{2q|q|}{\sqrt{ab}} \frac{R(\epsilon)}{\epsilon^2},
\end{align*}
and
\begin{align*}
- \frac{2q}{\epsilon} \Lambda (\epsilon) & = \frac{q}{\epsilon} \int_0^\omega {\rm d}\gamma \, \frac{\rho (\gamma)}{R(\gamma)}  \frac{R(\gamma) - R(\epsilon)}{\gamma - \epsilon} - \frac{2q|q|}{\sqrt{ab}} \frac{R(\epsilon)}{\epsilon^2} - \frac{2q^2}{\epsilon^2},
\end{align*}
hence
\begin{align*}
    \Lambda (\epsilon)^2 - \frac{2q}{\epsilon} \Lambda (\epsilon) & = \frac{\mathcal{J}}{4} - \frac{|q|}{\sqrt{ab}} \int_0^\omega {\rm d}\gamma \, \frac{\rho(\gamma)R(\epsilon)}{R(\gamma)} \frac{R(\gamma) - R(\epsilon)}{\epsilon(\gamma - \epsilon)}\\
    & \qquad \, + \frac{q^2}{ab} - \frac{q^2}{ab}\frac{(a+b)}{\epsilon}.
\end{align*}
On the RHS of \eqref{eq:ppip0prop1eq}, we first use \textbf{Lemma 1} and simplify the following term
\begin{align*}
\int_0^\omega {\rm d} \delta \, \rho(\delta)  \frac{\Lambda(\delta) - \Lambda(\epsilon)}{\delta - \epsilon} & = \frac{\mathcal{I}}{4} + \frac{|q|}{\sqrt{ab}} \int_0^\omega  {\rm d} \delta \, \rho(\delta)  \frac{R(\delta) - R(\epsilon)}{(\delta - \epsilon)\epsilon} \\
& \qquad  - \frac{|q|}{\sqrt{ab}} \int_0^\omega  {\rm d} \delta \, \rho(\delta)R(\delta) \frac{1}{\delta \epsilon} - q \int_0^\omega  {\rm d} \delta \, \rho(\delta) \frac{1}{\delta \epsilon}.
\end{align*}
Since
\begin{align*}
\frac{1}{\epsilon} \int_0^\omega {\rm d} \delta \, \rho(\delta)\Lambda(\delta) & = - \frac{\mathcal{K}}{4} +\frac{|q|}{\sqrt{ab}}\int_0^\omega {\rm d} \delta \, \rho(\delta) \frac{R(\delta)}{\delta \epsilon} + q \int_0^\omega {\rm d} \delta \, \frac{\rho(\delta)}{\delta \epsilon},
\end{align*}
we have
\begin{multline*}
\int_0^\omega {\rm d} \delta \, \rho (\delta) \frac{\Lambda (\epsilon) -\Lambda (\delta) }{\epsilon - \delta} + \frac{1}{\epsilon} \int_0^\omega  {\rm d} \delta \, \rho(\delta) \Lambda(\delta)\\
= \frac{\mathcal{I}}{4} - \frac{\mathcal{K}}{4} + \frac{|q|}{\sqrt{ab}} \int_0^\omega  {\rm d} \delta \, \rho(\delta)  \frac{R(\delta) - R(\epsilon)}{(\delta - \epsilon)\epsilon}.
\end{multline*}
Finally,
\begin{align*}
    &\, \Lambda (\epsilon)^2 - \frac{2q}{\epsilon} \Lambda (\epsilon) - \int_0^\omega {\rm d} \delta \, \rho (\delta) \frac{\Lambda (\epsilon) -\Lambda (\delta) }{\epsilon - \delta} - \frac{1}{\epsilon} \int_0^\omega  {\rm d} \delta \, \rho(\delta) \Lambda(\delta)\\
    = &\, \frac{\mathcal{J} - \mathcal{I}}{4} + \frac{\mathcal{K}}{4} - \frac{|q|}{\sqrt{ab}} \mathcal{O} + \frac{q^2}{ab} - \frac{q^2}{ab}\frac{(a+b)}{\epsilon}\\
    = &\, 0.
\end{align*}
\qed
\end{prop}
The integral approximation for \eqref{eq:ppip0rootstrans2} corresponding to the solution $\Lambda(\epsilon)$ given in \eqref{eq:ppip0prop1sol} is 
\begin{align} \nonumber
    t(\epsilon) & =  -G^{-1} - 2M + 2 \epsilon \Lambda(\epsilon)\\ \label{eq:ppip0tj2}
    & = -\int_0^\omega {\rm d}\delta \, \rho(\delta) \frac{\delta}{R(\delta)}\frac{R(\epsilon) - R(\delta)}{\epsilon - \delta}.
\end{align}
As a direct consequence of \textbf{Proposition \ref{prop1}}, we have the following result,
\begin{corollary} \label{cor2}
The function $t(\epsilon)$ defined in \eqref{eq:ppip0tj2} is a solution to the integral approximation of \eqref{eq:ppip0tj1},
\begin{equation} \label{eq:ppip0tj3}
    t(\epsilon)^2 = \frac{1}{G^2} + 2 \int_0^\omega {\rm d}\delta \, \rho(\delta) \frac{\epsilon (t(\epsilon) - t(\delta))}{\epsilon - \delta}.
\end{equation}
\end{corollary}
This particular solution $t(\epsilon)$ corresponds to the ground state since from \eqref{eq:egytj} in the continuum limit,
\begin{align} \nonumber
    E & = \frac{G}{2} \int_0^\omega {\rm d}\epsilon \, \rho(\epsilon) \epsilon t(\epsilon) +  \frac{1}{2}\int_0^\omega {\rm d}\epsilon \, \rho(\epsilon) \epsilon \\ \nonumber
    & = - \frac{G}{2} \int_0^\omega \int_0^\omega {\rm d}\epsilon \, {\rm d}\delta \, \rho(\epsilon) \rho(\delta) \frac{\epsilon \delta}{R(\delta)} \frac{R(\epsilon) - R(\delta)}{\epsilon - \delta} +  \frac{1}{2}\int_0^\omega {\rm d}\epsilon \, \rho(\epsilon) \epsilon
    \\ \label{eq:ppip0prop1sol2egy}
    & = - \frac{a + b}{4G} + |q|\sqrt{ab} - \frac{1}{2} \int_0^\omega {\rm d}\epsilon \, \rho(\epsilon) R(\epsilon) + \frac{1}{2}\int_0^\omega {\rm d}\epsilon \, \rho(\epsilon) \epsilon ,
\end{align}
the last step is due to expression $\mathcal{P}$ in \textbf{Lemma \ref{lemma1}}. The energy expression \eqref{eq:ppip0prop1sol2egy} coincides with the ground-state energy derived by mean-field analysis \cite{dilsz10} and integral approximation of both forms of Bethe Ansatz solutions discussed in Sect. \ref{sec:bae1cla} and \ref{sec:bae2cla} despite its limitations.
\section{The $p+ip$ model interacting with its environment} \label{sec:ppip1}

Consider the open model \eqref{eq:ppip1} with extra terms governed by parameter $\Gamma$,
$$ \mathcal{H} = H_0 + \Gamma \sum_{k=1}^L (S_k^+ + S_k^-). $$
\subsection{Bethe Ansatz equations and numerics}

The Bethe Ansatz solution for \eqref{eq:ppip1} was derived in \cite{lil16}: for each solution $\{v_j\}$ of the coupled equations
\begin{equation} \label{eq:ppip1bae}
\alpha + \sum_{k\neq j}^L \frac{2v_k}{v_k - v_j} + \sum_{l=1}^L \frac{z_l^2}{v_j-z_l^2} = -\frac{\beta^2}{v_j} \frac{\prod_{l=1}^L (1-v_j z_l^{-2})}{\prod_{k \neq j}^L (1-v_j v_k^{-1})},
\end{equation}
where $\alpha=1+\mathcal{G}^{-1}$, $\beta=\Gamma/\mathcal{G}$ and $j=1,\dots,L$,
there is a correspondence between $\{\mathfrak{t}_j\}$ in \eqref{eq:tj3} and $\{v_j \}$ via a change of variables 
\begin{equation} \label{eq:ppip1rootstrans}
    \mathfrak{t}_j = -\mathcal{G}^{-1} - 2L + 2z_j^2 \sum_{k=1}^L \frac{1}{z_j^2 - v_k}.
\end{equation}
This translates to the fact that \eqref{eq:tj3} and \eqref{eq:ppip1bae} are equivalent. It is worth mentioning that the difference between the closed and open model is that in the closed model, it is possible for the number of Bethe roots $M$ to be less than $L$, while in the open model we must have exactly $L$ Bethe roots \cite{l17, l18}. The energy is given by \eqref{eq:egytj},
\begin{align} \nonumber
    \mathcal{E} & = \frac{\mathcal{G}}{2} \sum_{j=1}^L z_j^2 \mathfrak{t}_j  + \frac{1}{2} \sum_{k=1}^L z_k^2 \\ \label{eq:ppip1baeegy}
    & = \mathcal{G}\alpha \sum_{k=1}^L v_k + \mathcal{G}\beta^2 \sum_{j=1}^L \frac{\prod_{l=1}^L (1-v_j z_l^{-2})}{\prod_{k \neq j}^L (1-v_j v_k^{-1})}. 
\end{align}
The corresponding eigenstate reads as \cite{cdv16}
$$ \prod_{j=1}^L \left( \frac{\Gamma}{v_j^2} + \mathcal{G} \sum_{k=1}^L \frac{z_k}{z_k^2 - v_j^2} S_k^\dagger \right) |0\rangle.  $$
\begin{table}
\caption{Numerical solutions for $\{v_k\}$ in \eqref{eq:ppip1bae} for various values of $\Gamma^2$ while $L=5$,  $\mathcal{G}=1$ and  $z_l^2=1,2,\dots,5$, are fixed. The energy is calculated according to \eqref{eq:ppip1baeegy}. The results are grouped corresponding to Fig. \ref{fig:ppip1baenumfig1}, \ref{fig:ppip1baenumfig2} and \ref{fig:ppip1baenumfig3} respectively.}
\small
\begin{tabular}{ |p{1.3cm}|p{9.25cm}|p{1.9cm}|  }
 \hline
 $\Gamma^2$ & Bethe roots & Energy\\
 \hline
 $10^{-6}$ & $(-1.5 \pm 1.3 \mi)\times 10^{-7}$, $-7 \times 10^{-7}$, $ -0.444 $, $-4.873$ & $-10.6343$ \\
 $2.3$ & $-0.122 \pm 0.363 \mi$, $ -1.104 $, $-1.484$, $-8.728$ & $-20.7818$\\
 $2.41305$ & $-0.122 \pm 0.380 \mi$, $-1.329$, $-1.336$, $-8.860$ & $-21.0716$\\
 $2.4131$ & $-0.122 \pm 0.380 \mi$, $-1.332 \pm 0.001 \mi $, $-8.860$ & $-21.0717$\\
 $2.42$ & $-0.121 \pm 0.381 \mi$, $-1.335 \pm 0.049 \mi $, $-8.868$ & $-21.0892$\\
 $5$ & $-0.028 \pm 0.712 \mi$, $-1.875 \pm 
 1.756 \mi$, $-12.846$ & $-26.5102$\\
 $8.1$ & $0.184 \pm 0.972 \mi$, $-1.235 \pm 3.560 \mi$, $-33.293$ & $-31.4424$\\
 \hline
 $8.5$ & $0.214 \pm 0.997 \mi$, $-1.073 \pm 3.731 \mi$, $-41.623$ & $-32.0068$\\
 $10.15$ & $0.339 \pm 1.080 \mi$, $-0.332 \pm 4.244 \mi$, $-1203.0$ & $-34.2096$\\
 $10.212$ & $0.344 \pm 1.082 \mi$, $-0.303 \pm 4.257 \mi$, $-70192$ & $-34.2889$\\
 $10.213$ & $0.344 \pm 1.082 \mi$, $-0.303 \pm 4.258 \mi$, $-907423$ & $-34.2901$\\
 $10.2132$ & $0.344 \pm 1.082 \mi$, $-0.303 \pm 4.258 \mi$, $654976$ & $-34.2904$\\
 $10.5$ & $0.366 \pm 1.094 \mi$, $-0.169 \pm 4.315 \mi$, $268.233$ & $-34.6541$\\
 $13$ & $0.552 \pm 1.167 \mi$, $0.938 \pm 4.523 \mi$, $30.566$ & $-37.6403$\\
 \hline
 $14$ & $0.623 \pm 1.184 \mi$, $1.326 \pm 4.493 \mi$, $23.411$ & $-38.7557$\\
 $20$ & $1.002 \pm 1.193 \mi$, $2.888 \pm 3.771 \mi$, $11.2682$ & $-44.7606$\\
 $48.5$ & $1.977 \pm 0.223 \mi$, $4.019 \pm 0.709 \mi$, $5.592$ & $-65.3902$\\
 $49.455$ & $1.9999 \pm 0.016 \mi$, $4.015 \pm 0.600 \mi$, $5.541$ & $-65.9570$\\
 $49.4633$ & $1.9871$, $2.0130$, $4.015 \pm 0.599 \mi$, $5.541$ & $-65.9619$\\
 $51.7$ & $1.708$, $2.400$, $4.002 \pm 0.207 \mi$, $5.430$ & $-67.2686$\\
 $52$ & $1.6925$, $2.4310$, $4.000 \pm 0.040 \mi$, $5.416$ & $-67.4417$\\
 $52.013$ & $1.692$, $2.432$, $3.987$, $4.013$, $5.415$ & $-67.4492$\\
 $54.5$ & $1.597$, $2.674$, $3.381$, $4.558$, $5.300$ & $-68.8656$\\
 $56.34$ & $1.548$, $2.988$, $3.012$, $4.727$, $5.213$ & $-69.8931$\\
 $56.3443$ & $1.548$, $3.000 \pm 0.009 \mi$, $4.727$, $5.213$ & $-69.8954$\\
 $58.925$ &  $1.496$, $2.982 \pm 0.319 \mi$, $4.986$, $5.014$ & $-71.3090$\\
 $58.96$ & $1.496$, $2.982 \pm 0.321 \mi$, $5.000 \pm 0.019 \mi$ & $-71.3279$\\
 $66$ & $1.402$, $2.970 \pm 0.450 \mi$, $5.021 \pm 0.203 \mi$ & $-75.0345$\\
 $77.59$ & $1.316$, $2.981 \pm 0.424 \mi$, $5.000 \pm 0.013 \mi$ & $-80.7337$\\
 $77.64$ & $1.316$, $2.981 \pm 0.423 \mi$, $4.991$, $5.009$ & $-80.7573$\\
 $103.565$ & $1.223$, $3.000 \pm 0.006 \mi$, $4.589$, $5.272$ & $-92.1382$\\
 $105$ & $1.220$, $2.905$, $3.096$, $4.579$, $5.276$ & $-92.7243$\\
 $250$ & $1.101$, $2.325$, $3.420$, $4.367$, $5.301$ & $-139.209$\\
 $10^3$ & $1.039$, $2.119$, $3.201$, $4.244$, $5.229$ & $-271.579$\\
 $10^7$ & $1.000$, $2.001$, $3.002$, $4.003$, $5.004$ & $-26513.6$\\
 \hline
\end{tabular}
\label{tab:ppip1baenum}
\end{table}
\normalsize
\begin{figure}
    \begin{subfigure}{0.49\textwidth}
    \includegraphics[scale=0.3]{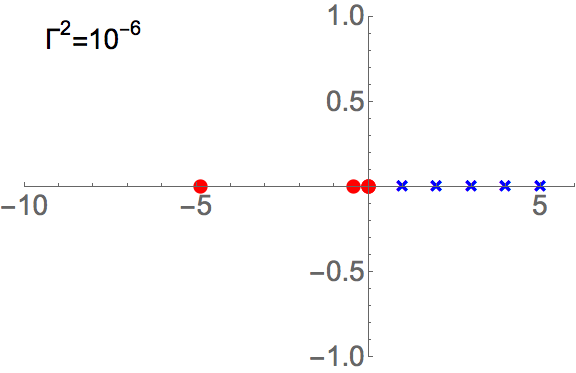}
    \caption{}
    \end{subfigure}
    \begin{subfigure}{0.49\textwidth}
    \includegraphics[scale=0.3]{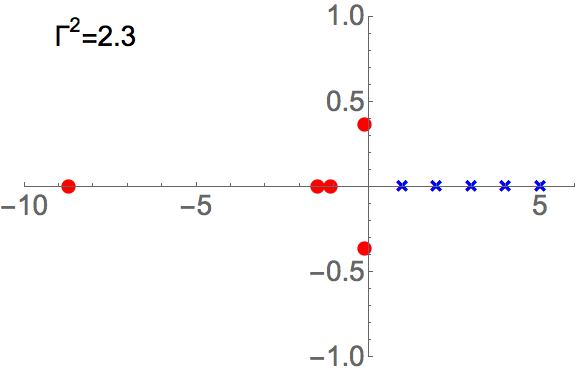}
    \caption{}
    \end{subfigure}\\
    \begin{subfigure}{0.49\textwidth}
    \includegraphics[scale=0.3]{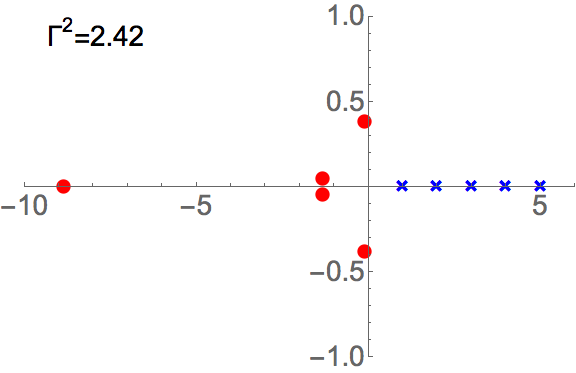}
    \caption{}
    \end{subfigure}
    \begin{subfigure}{0.49\textwidth}
    \includegraphics[scale=0.3]{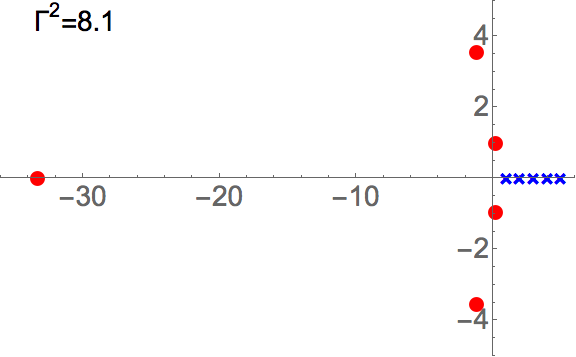}
    \caption{}
    \end{subfigure}
\caption{The Bethe roots $\{v_k\}$ are depicted by solid dots. The crosses mark positions $z_l^2=1,2,3,4,5$. In (a), three roots are coincident at $0$. As $\Gamma^2$ increases, the three coincident roots depart from $0$. From (b) to (c), 2 real roots meet for a value of $\Gamma^2$ between $2.3$ and $2.42$. After the two roots meet, they separate and form a complex-conjugate pair. In (d), the negative real root diverges as $\Gamma^2$ continues to increase.\\}
\label{fig:ppip1baenumfig1}
    \begin{subfigure}{0.49\textwidth}
    \includegraphics[scale=0.3]{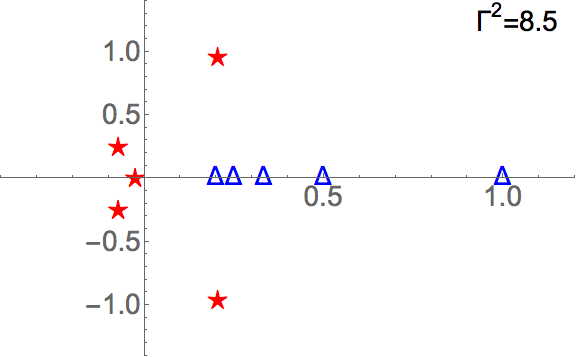}
    \caption{}
    \end{subfigure}
    \begin{subfigure}{0.49\textwidth}
    \includegraphics[scale=0.3]{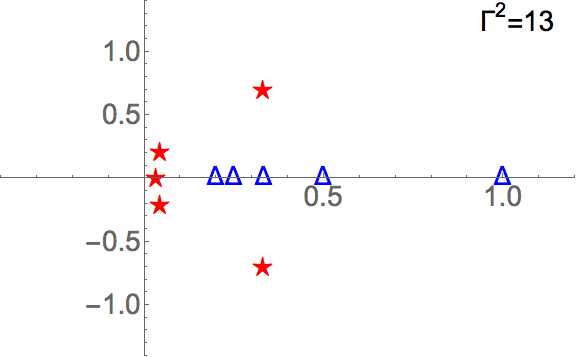}
    \caption{}
    \end{subfigure}
\caption{The inverted roots $\{v_k^{-1} \in \mathbb{C}\}$ are depicted by stars to visualize the diverging real root. The triangles are the $\{z_l^{-2}\}$. From (a) to (b), the inverted real root traverses $0$ and becomes positive.}
\label{fig:ppip1baenumfig2}
\end{figure}
\begin{figure}
    \begin{subfigure}{0.49\textwidth}
    \includegraphics[scale=0.3]{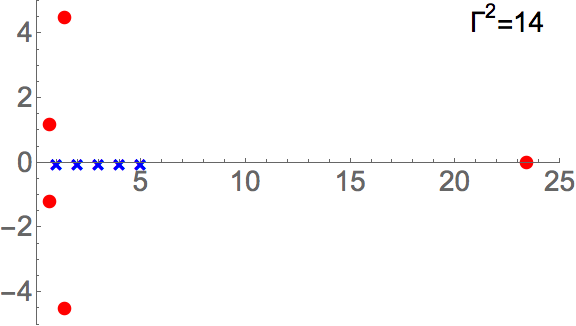}
    \caption{}
    \end{subfigure}
    \begin{subfigure}{0.49\textwidth}
    \includegraphics[scale=0.3]{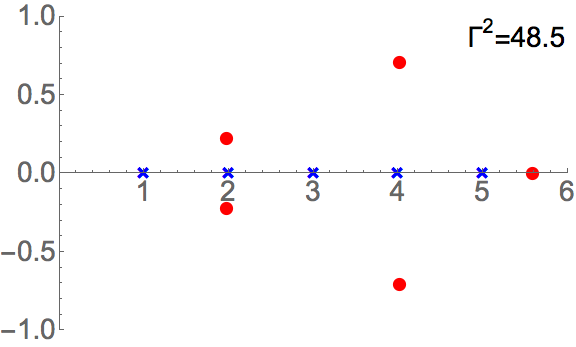}
    \caption{}
    \end{subfigure}\\
    \begin{subfigure}{0.49\textwidth}
    \includegraphics[scale=0.3]{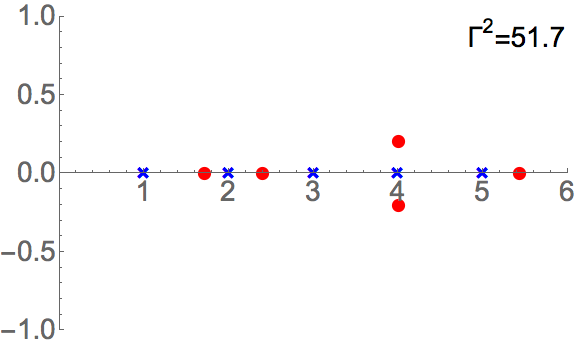}
    \caption{}
    \end{subfigure}
    \begin{subfigure}{0.49\textwidth}
    \includegraphics[scale=0.3]{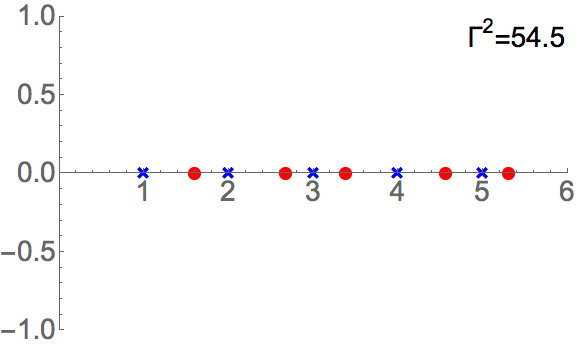}
    \caption{}
    \end{subfigure}\\
    \begin{subfigure}{0.49\textwidth}
    \includegraphics[scale=0.3]{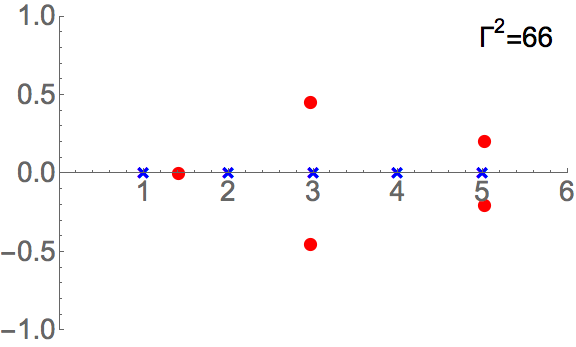}
    \caption{}
    \end{subfigure}
    \begin{subfigure}{0.49\textwidth}
    \includegraphics[scale=0.3]{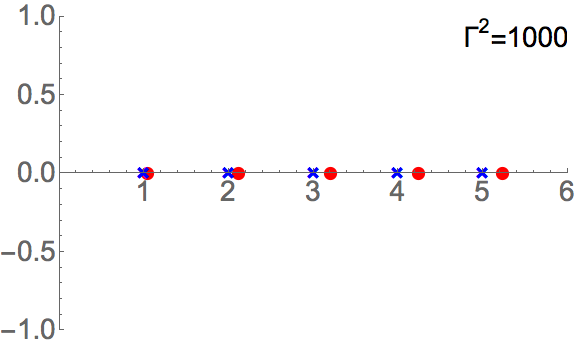}
    \caption{}
    \end{subfigure}
\caption{In (a), the real root decreases from $+\infty$ and approaches 5 as $\Gamma^2$ continues to increase. From (b) to (c), two complex-conjugate roots meet at position $2$ and become real. From (c) to (d), two complex-conjugate roots meet at position $4$ and become real. From (d) to (e), two real roots meet and separate as a complex-conjugate pair at position $3$, and then two other real roots behave similarly at position $5$. In (f) as $\Gamma^2$ becomes large, all roots approach $\{z_l^2\}$ from right.}
\label{fig:ppip1baenumfig3}
\end{figure}

We perform some numerical analysis to study the behaviour of the Bethe roots $\{v_j\}$. We consider a small-sized case where $L=5$. As we send $\Gamma$ from $0$ to a large number, numerical results (see Tab. \ref{tab:ppip1baenum}) show that one of $\{v_j\}$ in \eqref{eq:ppip1bae} diverges around parameters $(\mathcal{G}=1, \Gamma^2 = 10.213)$. Also at certain values of $(\mathcal{G}, \Gamma)$, we have two real roots meeting to form a complex-conjugate pair and vice versa, see Fig.\phantom{ }\ref{fig:ppip1baenumfig1}, \ref{fig:ppip1baenumfig2} and \ref{fig:ppip1baenumfig3}.
\subsection{Conserved operator eigenvalue method for the open model}

The numerical analysis of the distribution of the Bethe roots gives no clear indications of a solution curve for a large particle number. However, since \eqref{eq:ppip1bae} is invariant under complex conjugation, again assuming the absence of degeneracy in the spectrum of the set of conserved operators, each solution set $\{ v_j \}$ consists of complex-conjugate pairs or real numbers. Hence from \eqref{eq:ppip1rootstrans}, it is clear to see that all solution sets of $\{\mathfrak{t}_j\}$ are real. We now consider the integral approximation for \eqref{eq:tj3}, which reads as
\begin{align}\label{eq:tj5}
    \mathfrak{t}(\epsilon)^2 & = \frac{1}{\mathcal{G}^2} + \frac{F^2 \epsilon^{-1}}{\mathcal{G}^2} + 2\int_0^\omega {\rm d}\delta \, \rho(\delta) \frac{\delta}{\epsilon - \delta} \big(\mathfrak{t}(\epsilon) - \mathfrak{t}(\delta)\big),
\end{align}
where $F = - 2\Gamma$. Our task is to find a solution to \eqref{eq:tj5} that corresponds to the ground state. We resort to mean-field analysis for suggestions of a possible solution. The mean-field analysis for the open model and its results are included in Appendix A. We consider the operators $\mathcal{T}_j$ defined in \eqref{eq:tj1}. From \eqref{eq:ppip1opmean1} and \eqref{eq:ppip1opmean2}, we derive the mean-field expression for $\mathcal{T}_j$ as
\begin{align} \label{eq:tj4}
\langle \mathcal{T}_j \rangle & = -\frac{z_j^2 + F(\Delta + F)}{\mathcal{G}\mathcal{R}(z_j^2)} + \sum_{k\neq j}^L \frac{1}{z_k^2 - z_j^2} \left( \frac{z_j^2 \mathcal{R}(z_k^2)}{\mathcal{R}(z_j^2) } - z_k^2 \right), \quad j=1,\dots,L,
\end{align}
where
\begin{align*}
\mathcal{R}(z) & = \sqrt{z^2 + z (\Delta+F)^2}, \qquad \Delta \in \mathbb{R}.
\end{align*}
The integral approximation for \eqref{eq:tj4} is
\begin{align} \label{eq:tj6}
    \mathcal{T} (\epsilon) & = -\frac{\epsilon + F(\Delta + F)}{\mathcal{G}\mathcal{R}(\epsilon)} + \int_0^\omega {\rm d}\delta \, \rho(\delta) \frac{1}{\delta - \epsilon} \left( \frac{\epsilon \mathcal{R}(\delta)}{\mathcal{R}(\epsilon) }- \delta \right).
\end{align}
We will establish that \eqref{eq:tj6} is in fact a solution to \eqref{eq:tj5}, i.e.
$$ \mathfrak{t} (\epsilon) = \mathcal{T} (\epsilon),$$
with the ``gap'' equation determining the value of $\Delta$,
\begin{align} \label{eq:ppip1gap4}
    \frac{\Delta}{\mathcal{G}(\Delta + F)} = \int_0^\omega {\rm d}\epsilon \, \rho(\epsilon) \frac{\epsilon}{\mathcal{R}(\epsilon)}.
\end{align}
In addition, this solution $\mathfrak{t}(\epsilon)$ corresponds to the ground state. We again establish some useful results to assist our proof.
\begin{lemma} \label{lemma2}
Given equation \eqref{eq:ppip1gap4}, let $$T(\epsilon) = -\int_0^\omega {\rm d}\delta \, \rho(\delta) \frac{\delta}{\mathcal{R}(\delta)}\frac{\mathcal{R}(\epsilon) - \mathcal{R}(\delta)}{\epsilon - \delta}, \quad \mathcal{C} = \int_0^\omega {\rm d}\epsilon \, \frac{\rho(\epsilon)}{\mathcal{R}(\epsilon)}, $$
then 
\begin{align} \label{eq:ppip1tj1}
T(\epsilon)^2 & = \frac{\Delta^2}{\mathcal{G}^2(\Delta+F)^2} + 2 \int_0^\omega {\rm d}\delta \, \rho(\delta) \frac{\delta (T(\epsilon) - T(\delta))}{\epsilon - \delta},\\ \label{eq:ppip1tj2}
T(\epsilon) & = \mathcal{C} \mathcal{R}(\epsilon) - L - \epsilon \int_0^\omega {\rm d}\delta \, \rho(\delta) \frac{1}{\mathcal{R}(\delta)}\frac{\mathcal{R}(\epsilon) - \mathcal{R}(\delta)}{\epsilon - \delta},\\ \label{eq:ppip1tj3}
\mathfrak{t} (\epsilon) & = - \frac{F}{\mathcal{G}(\Delta + F)} \frac{\mathcal{R}(\epsilon)}{\epsilon} + T(\epsilon).
\end{align}
Proof:
In \eqref{eq:ppip0claconstraints3}, set $$\frac{1}{G} = \frac{\Delta}{\mathcal{G}(\Delta + F)} ,\quad q = b = 0 ,\quad a = -(\Delta + F)^2, $$
then we have $R(\epsilon) = \mathcal{R}(\epsilon) $ and $C = \mathcal{C}$. Hence from \eqref{eq:ppip0tj2} and by \textbf{Corollary \ref{cor2}},
$T(\epsilon) = t(\epsilon)$ satisfying \eqref{eq:ppip0tj3},
\begin{equation*}
    T(\epsilon)^2 =  \frac{\Delta^2}{\mathcal{G}^2(\Delta+F)^2} + 2 \int_0^\omega {\rm d}\delta \, \rho(\delta) \frac{\delta (T(\epsilon) - T(\delta))}{\epsilon - \delta}.
\end{equation*}
Next
\begin{align*}
    T(\epsilon) & = -\int_0^\omega {\rm d}\delta \, \rho(\delta) \frac{\delta - \epsilon + \epsilon}{\mathcal{R}(\delta)}\frac{\mathcal{R}(\epsilon) - \mathcal{R}(\delta)}{\epsilon - \delta}\\
    & = \mathcal{C} \mathcal{R}(\epsilon) - L - \epsilon \int_0^\omega {\rm d}\delta \, \rho(\delta) \frac{1}{\mathcal{R}(\delta)}\frac{\mathcal{R}(\epsilon) - \mathcal{R}(\delta)}{\epsilon - \delta}.
\end{align*}
Finally,
\begin{align*} 
\mathfrak{t} (\epsilon) & = -\frac{\epsilon + F(\Delta + F)}{\mathcal{GR}(\epsilon)} + \int_0^\omega {\rm d}\delta \, \rho(\delta) \frac{1}{\delta - \epsilon} \left( \frac{\epsilon \mathcal{R}(\delta)}{\mathcal{R}(\epsilon)} - \epsilon - \delta + \epsilon \right)\\
& = - \frac{F}{\mathcal{G}(\Delta + F)} \frac{\mathcal{R}(\epsilon)}{\epsilon} + \mathcal{C} \mathcal{R}(\epsilon) - L - \epsilon \int_0^\omega {\rm d}\delta \, \rho(\delta) \frac{1}{\mathcal{R}(\delta)} \frac{\mathcal{R}(\delta) - \mathcal{R}(\epsilon)}{\delta - \epsilon}.
\end{align*}
\qed
\end{lemma}
Now we show the following:
\begin{prop} \label{prop2}
The following function
\begin{equation}
\mathfrak{t}(\epsilon) = - \frac{F}{\mathcal{G}(\Delta + F)} \frac{\mathcal{R}(\epsilon)}{\epsilon} - \int_0^\omega {\rm d}\delta \, \rho(\delta) \frac{\delta}{\mathcal{R}(\delta)}\frac{\mathcal{R}(\epsilon) - \mathcal{R}(\delta)}{\epsilon - \delta} 
\end{equation}
is a solution to the integral equation \eqref{eq:tj5} with $\Delta$ subject to \eqref{eq:ppip1gap4}. \\
Proof: We adopt the notation introduced in \textbf{Lemma \ref{lemma2}}, rewriting
$$ \mathfrak{t}(\epsilon) = - \frac{F}{\mathcal{G}(\Delta + F)} \frac{\mathcal{R}(\epsilon)}{\epsilon} + T(\epsilon).$$
Since
\begin{align*}
    \mathfrak{t}(\epsilon)^2 & = \frac{F^2}{\mathcal{G}^2 (\Delta+F)^2} + \frac{F^2 \epsilon^{-1}}{\mathcal{G}^2} - \frac{2F}{\mathcal{G} (\Delta+F)} \frac{\mathcal{R}(\epsilon)}{\epsilon} T(\epsilon) + T(\epsilon)^2,
\end{align*}
and
\begin{align*}
    2\int_0^\omega {\rm d}\delta \, \rho(\delta) \frac{\delta}{\epsilon - \delta} \big(\mathfrak{t}(\epsilon) - \mathfrak{t}(\delta)\big) & = - \frac{2F}{\mathcal{G} (\Delta+F)} \int_0^\omega {\rm d}\delta \,  \frac{\rho(\delta)\delta}{\epsilon - \delta} \left( \frac{\mathcal{R}(\epsilon)}{\epsilon} - \frac{\mathcal{R}(\delta)}{\delta} \right)\\
    & \qquad + 2\int_0^\omega {\rm d}\delta \, \rho(\delta) \frac{\delta}{\epsilon - \delta} (T(\epsilon) - T(\delta)),
\end{align*}
then from \eqref{eq:ppip1tj1} we are able to simplify equation \eqref{eq:tj5} as
\begin{align} \nonumber
    & -\frac{F^2 + \Delta^2}{\mathcal{G}^2 (\Delta+F)^2} + \frac{1}{\mathcal{G}^2} \\ \label{eq:ppip1prop2temp1}
    & \qquad + \frac{2F}{\mathcal{G}(\Delta + F)} \left( \frac{\mathcal{R}(\epsilon)}{\epsilon} T(\epsilon) - \int_0^\omega {\rm d}\epsilon \,  \frac{\rho(\delta)\delta}{\epsilon - \delta} \left( \frac{\mathcal{R}(\epsilon)}{\epsilon} - \frac{\mathcal{R}(\delta)}{\delta} \right) \right) = 0.
\end{align}
Since
\begin{align*}
    \int_0^\omega {\rm d}\delta \,  \frac{\rho(\delta)\delta}{\epsilon - \delta} \left( \frac{\mathcal{R}(\epsilon)}{\epsilon} - \frac{\mathcal{R}(\delta)}{\delta} \right) & =  \int_0^\omega {\rm d}\delta \, \rho(\delta)\left( \frac{\mathcal{R}(\epsilon) - \mathcal{R}(\delta)}{\epsilon - \delta} - \frac{\mathcal{R}(\epsilon)}{\epsilon} \right),
\end{align*}
\begin{align*}
    \frac{\mathcal{R}(\epsilon)}{\epsilon} T(\epsilon) & = \mathcal{C} \frac{\epsilon^2 + \epsilon(\Delta+F)^2}{\epsilon} - L \frac{\mathcal{R}(\epsilon)}{\epsilon} - \mathcal{R}(\epsilon) \int_0^\omega {\rm d}\delta \,  \frac{\rho(\delta)}{\mathcal{R}(\delta)}\frac{\mathcal{R}(\epsilon) - \mathcal{R}(\delta)}{\epsilon - \delta},
\end{align*}
and we use the special case of $q=b=0$ for expression $\mathcal{O}$ in \textbf{Lemma \ref{lemma1}}, where $ (2|q|/\sqrt{ab})$, $(a+b)$ and $G^{-1}$ are replaced by $\mathcal{C}$, $ -(\Delta+F)^2$ and $\Delta \mathcal{G}^{-1} (\Delta+F)^{-1}$ respectively, consequently we have
\begin{multline*}
    - \mathcal{R}(\epsilon) \int_0^\omega {\rm d}\delta \,  \frac{\rho(\delta)}{\mathcal{R}(\delta)}\frac{\mathcal{R}(\epsilon) - \mathcal{R}(\delta)}{\epsilon - \delta} - \int_0^\omega {\rm d}\delta \, \rho(\delta) \frac{\mathcal{R}(\epsilon) - \mathcal{R}(\delta)}{\epsilon - \delta} \\
    = \, - \frac{\Delta}{\mathcal{G}(\Delta+F)} - \mathcal{C} (\Delta + F)^2 - \mathcal{C}\epsilon.
\end{multline*}
Hence
\begin{align*}
    \frac{\mathcal{R}(\epsilon)}{\epsilon} T(\epsilon) - \int_0^\omega {\rm d}\delta \,  \frac{\rho(\delta)\delta}{\epsilon - \delta} \left( \frac{\mathcal{R}(\epsilon)}{\epsilon} - \frac{\mathcal{R}(\delta)}{\delta} \right) & = - \frac{\Delta}{\mathcal{G}(\Delta + F)}.
\end{align*}
Now the LHS of \eqref{eq:ppip1prop2temp1} is reduced to the following
\begin{equation*}
    -\frac{F^2 + \Delta^2}{\mathcal{G}^2 (\Delta+F)^2} + \frac{1}{\mathcal{G}^2} - \frac{2F\Delta}{\mathcal{G}^2 (\Delta+F)^2} = 0.
\end{equation*}
\qed
\end{prop}
We derive the energy expression using $\mathfrak{t}(\epsilon)$ from \eqref{eq:egytj},
\begin{align*}
    \mathcal{E} & = \frac{\mathcal{G}}{2} \int_0^\omega {\rm d}\epsilon \, \rho(\epsilon) \epsilon \mathfrak{t} (\epsilon) + \frac{1}{2} \int_0^\omega {\rm d}\epsilon \, \rho(\epsilon) \epsilon \\
    & = - \frac{F}{2(\Delta + F)} \int_0^\omega {\rm d} \epsilon \, \rho(\epsilon) \mathcal{R}(\epsilon) + \frac{1}{2} \int_0^\omega {\rm d}\epsilon \, \rho(\epsilon) \epsilon \\
    & \qquad \qquad \qquad \qquad \qquad - \frac{\mathcal{G}}{2} \int_0^\omega \int_0^\omega {\rm d} \epsilon \, {\rm d} \delta \, \rho(\epsilon) \rho(\delta) \frac{\epsilon \delta \big( \mathcal{R}(\epsilon) - \mathcal{R}(\delta) \big)}{\mathcal{R}(\delta) (\epsilon - \delta)}.
\end{align*}
Again from the special case for expression $\mathcal{P}$ in \textbf{Lemma \ref{lemma1}}, where $b$, $a$ and $G^{-1}$ are replaced by $0$, $ -(\Delta+F)^2$ and $\Delta \mathcal{G}^{-1} (\Delta+F)^{-1}$ respectively, we have
\begin{multline*}
    - \frac{\mathcal{G}}{2} \int_0^\omega \int_0^\omega {\rm d} \epsilon \, {\rm d} \delta \, \rho(\epsilon) \rho(\delta) \frac{\epsilon \delta \big( \mathcal{R}(\epsilon) - \mathcal{R}(\delta) \big)}{\mathcal{R}(\delta) (\epsilon - \delta)}\\
    = \frac{\Delta^2}{4\mathcal{G}} - \frac{\Delta}{2(\Delta + F)} \int_0^\omega {\rm d} \epsilon \, \rho(\epsilon) \mathcal{R}(\epsilon).
\end{multline*}
Hence
\begin{align*}
    \mathcal{E} & = \frac{\Delta^2}{4\mathcal{G}} + \frac{1}{2} \int_0^\omega {\rm d}\epsilon \, \rho(\epsilon) \epsilon - \frac{1}{2} \int_0^\omega {\rm d} \epsilon \, \rho(\epsilon) \mathcal{R}(\epsilon).
\end{align*}
This energy expression is consistent with \eqref{eq:ppip1egy} (see Appendix A) for the ground state, which is derived from mean-field analysis.
\section{Conclusion} \label{sec:conclusion}

The continuum limit approximation for calculating the ground-state energy of the 
$p+ip$ model, via the Bethe Ansatz solution, was studied.  Starting with the closed model, we revisited the formulations of \cite{dilsz10,rdo10,admor02} which undertake calculations by assuming a form of density function for the Bethe root distribution. It was found that this approach does not provide a consistent solution for particular choices of the momentum density distribution. This was established by a close examination of the case known as the Moore-Read line, where it is known that all the ground-state Bethe roots collapse at the origin \cite{ilsz09,dilsz10,rdo10,vdv14}. An alternative approach, which avoids the need to postulate a form for the Bethe root density,  was proposed  in terms of the coupled equations satisfied by the conserved operator eigenvalues \cite{cdvv15,cvd17}. In this case a solution corresponding to the ground state in the continuum limit was found. Curiously, the expression obtained for the ground-state energy coincides with that obtained by the Bethe root distribution.  That is to say that although the Bethe root approach involves a flawed methodology, it nonetheless produces the correct answer! In both cases the ground-state energy per particle is exactly the same as the mean-field prediction in the limit of infinite particle number.

The conserved operator eigenvalue approach was then extended to accommodate the open case based on results from \cite{l17,cdv16,l18}. Again, it was found that the result for the ground-state energy is in agreement with mean-field calculations. The open case can be considered as a model allowing for the exchange of particle between the system and the environment.  It is important to note that in this case there is no signature of any quantum phase transition, no matter how weak the environment coupling is, which is in stark contrast to the closed system. A similar scenario was considered in \cite{dilz11} where the environment was modelled by a single bosonic degree of freedom in such a way that integrability was preserved. There too, arbitrarily small coupling to the environment was found to annihilate the existence of any quantum phase transition.      

 For future work it would be natural to extend this analysis to calculate the leading order finite-size correction to the ground-state energy, for both open and closed models. For the closed $s$-wave pairing systems there are results obtained by series expansions \cite{r77}, which can be extrapolated and shown to be valid more generally \cite{pc11}.    
Obtaining analogous expressions for the open and closed $p+ip$-pairing systems appears to be entirely feasible.

\section*{Acknowledgements}
\noindent
This work was supported by the Australian Research Council through Discovery Project DP150101294.

\begin{appendices}
\section{Mean-field analysis for the open model}
\noindent 
We introduce the following notation adopted in \cite{lmm15}:
$$ Q^\dagger = \sum_{k=1}^L z_k S_k^+, \, Q = \sum_{k=1}^L z_k S_k^-. $$ Then let $\hat{\Delta} =2\mathcal{G}\langle Q \rangle $, $\hat{\Delta}^\dagger =2\mathcal{G}\langle Q^\dagger \rangle $ and $\Delta = |\hat{\Delta}| $. The extended Hamiltonian \eqref{eq:ppip1} can be rewritten as
\begin{align*} 
    \mathcal{H} & \approx \sum_{k=1}^L z_k^2 S_k^z - \mathcal{G}Q\langle Q^\dagger \rangle - \mathcal{G}Q^\dagger \langle Q \rangle + \mathcal{G} \langle Q^\dagger \rangle \langle Q \rangle  + \mu \left( \sum_{k=1}^L \left\langle  S_k^z \right\rangle - \sum_{k=1}^L S_k^z\right)\\
    & \qquad - \frac{F}{2} Q^\dagger - \frac{F}{2} Q +  \frac{1}{2}\sum_{k=1}^L z_k^2 \\
    & = \frac{\Delta^2}{4\mathcal{G}} + \frac{1}{2}\sum_{ k=1 }^L z_k^2 + \sum_{k=1}^L \begin{pmatrix}
    z_k^2/2 & - (\hat{\Delta}+F) z_k/2 \\ \nonumber
    -(\hat{\Delta}^\dagger+F) z_k/2 & -z_k^2/2
    \end{pmatrix}.
\end{align*}
Note that in the mean-field approximation for this extended model, the Lagrange multiplier is $\mu = 0$. The matrix form is derived from the representation of $\mathcal{H}$ acting on $(\mathbb{C}^2) ^ {\otimes L}$. Consider the following eigenvalue problem,
$$\begin{pmatrix}
        z_k^2/2 & - (\hat{\Delta} + F) z_k/2 \\
        -(\hat{\Delta}^\dagger + F) z_k/2 & -z_k^2/2
    \end{pmatrix}
    \begin{pmatrix}
    v_k\\
    u_k
    \end{pmatrix}
     = \lambda_k
    \begin{pmatrix}
    v_k\\
    u_k
    \end{pmatrix}, \quad k=1,\dots,L,$$
by minimizing each eigenvalue, we derive the ground-state energy
\begin{equation} \label{eq:ppip1egy}
     \mathcal{E}_0 = \frac{\Delta^2}{4\mathcal{G}} + \frac{1}{2}\sum_{ k=1 }^L z_k^2 - \frac{1}{2} \sum_{k=1}^L \sqrt{z_k^4 + \chi^2 z_k^2},
\end{equation}
where $\chi = |\hat{\chi}|$, $\hat{\chi} = \hat{\Delta} + F$. Then we calculate the following 
\begin{align*}
    \langle S_k^z \rangle & = \frac{1}{2} (|v_k|^2 - |u_k|^2 ) = -\frac{1}{2} \frac{z_k^2}{\sqrt{z_k^4 + \chi^2 z_k^2}},\\
    \langle S_k^+ \rangle & = u_k v_k^* = \frac{\hat{\chi}^\dagger z_k}{2} \frac{1}{\sqrt{z_k^4 + \chi^2 z_k^2}},\\
    \langle S_k^+ \rangle & = u_k^* v_k = \frac{\hat{\chi} z_k}{2} \frac{1}{\sqrt{z_k^4 + \chi^2 z_k^2}}.
\end{align*}
Apply Hellmann-Feynman theorem, since
\begin{align*}
    \left\langle  \frac{\partial \mathcal{H}}{\partial \mathcal{G}} \right\rangle = - \frac{\Delta^2}{4\mathcal{G}^2}, \quad 
    \frac{\partial \mathcal{E}_0}{\partial \mathcal{G}} = \frac{\Delta^2}{4\mathcal{G}^2} - \frac{1}{2} \sum_{k=1}^L \frac{\chi z_k^2}{\sqrt{z_k^4 + \chi^2 z_k^2}} \frac{\partial \chi}{\partial \mathcal{G}},
\end{align*}
and
\begin{align} \label{eq:nabladelta}
\chi^2 & = \Delta^2 +F (\hat{\Delta} + \hat{\Delta}^\dagger) + F^2,\\ \nonumber
\frac{\partial \chi}{\partial \mathcal{G}} & = \frac{\Delta^2}{\mathcal{G} \chi} + \frac{F}{\mathcal{G} \chi} {\rm Re} \, \hat{\Delta},
\end{align}
we have 
\begin{equation} \label{eq:ppip1gap1}
     \sum_{k = 1}^L \frac{z_k^2}{\sqrt{z_k^4 + \chi^2 z_k^2}} = \frac{1}{\mathcal{G}} \frac{\Delta^2}{\Delta^2 + F {\rm Re} \, \hat{\Delta}}.
\end{equation}
Furthermore from \eqref{eq:nabladelta},
\begin{align*}
    \frac{\partial \chi}{\partial F} & = \frac{{\rm Re} \, \hat{\Delta}}{\chi} +  \frac{F}{\chi},
\end{align*}
and
\begin{align*}
    \left\langle \frac{\partial \mathcal{H}}{\partial F} \right\rangle & = -\frac{1}{2}\left( \frac{\hat{\Delta}^*}{2\mathcal{G}} +\frac{\hat{\Delta}}{2\mathcal{G}} \right) = -\frac{{\rm Re} \, \hat{\Delta}}{2\mathcal{G}},\\
    \frac{\partial \mathcal{E}_0}{\partial F} & = - \frac{1}{2} \sum_{k =1}^L \frac{z_k^2}{\sqrt{z_k^4 + \chi^2 z_k^2}} ({\rm Re} \, \hat{\Delta}  + F),
\end{align*}
hence
\begin{equation} \label{eq:ppip1gap2}
    \sum_{k = 1}^L \frac{z_k^2}{\sqrt{z_k^4 + \chi^2 z_k^2}} = \frac{1}{\mathcal{G}} \frac{{\rm Re} \, \hat{\Delta}}{{\rm Re} \, \hat{\Delta} + F}.
\end{equation}
Comparing \eqref{eq:ppip1gap1} and \eqref{eq:ppip1gap2}, and assuming $F \neq 0$, we conclude that
\begin{align} \label{eq:ppip1gapreal}
    {\rm Im}\, \hat{\Delta} & = 0,
\end{align}
i.e. $ \hat{\Delta} = \hat{\Delta}^\dagger = \Delta $. It follows that $\chi = \hat{\chi} = \hat{\chi}^\dagger = \Delta + F$. Hence we immediately have the following results
\begin{align} \label{eq:ppip1opmean1}
\langle S_k^z \rangle & = -\frac{1}{2} \frac{z_k^2}{\sqrt{z_k^4 + (\Delta + F)^2 z_k^2}}, \\ \label{eq:ppip1opmean2} 
\langle S_k^+ \rangle & = \langle S_k^- \rangle = \frac{1}{2} \frac{(\Delta + F) z_k}{\sqrt{z_k^4 + (\Delta + F)^2 z_k^2}},
\end{align}
and the ``gap" equation
\begin{equation} \label{eq:ppip1gap3}
    \frac{\Delta}{\mathcal{G}(\Delta + F)} = \sum_{k=1}^L \frac{z_k^2}{\sqrt{z_k^4 + (\Delta + F)^2 z_k^2}}. 
\end{equation}

\end{appendices}


\end{document}